\documentclass[a4paper,11pt]{scrreprt}
\usepackage[language=english,fpms,emblem]{umonsCover}

\umonsAuthor{ Ludovic \textsc{Pirard} }

\umonsTitle{Spatial Audio and Individualized HRTFs using a Convolutional Neural Network (CNN)}
\umonsSubtitle{ISIA Departement}
\umonsDocumentType{Master Thesis for the Master’s Degree in Electrical Engineering with a spe-
cialization in Artificial Intelligence and Smart Communication}

\umonsSupervisor{Promoter : \textsc{Thierry Dutoit} \\ Co-Promoter : Loïc \textsc{Reboursière}}

\umonsDate{Academic year 2022-2023}

\usepackage{graphicx}
\usepackage{subcaption}
\usepackage{nomencl}
\usepackage[nottoc, notlot, notlof]{tocbibind}
\usepackage[perpage]{footmisc}
\graphicspath{{image/}}

\usepackage{siunitx}

\usepackage{hyperref}

\usepackage{amsmath} 
\usepackage{cleveref}
\usepackage{graphicx}
\usepackage{comment}
\usepackage[utf8]{inputenc}

\begin{document}
\umonsCoverPage
\pagenumbering{roman}

\chapter*{Abstract}

Spatial audio and 3-Dimensional sound rendering techniques play a pivotal and essential role in immersive audio experiences. Head-Related Transfer Functions (HRTFs) are acoustic filters which represent how sound interacts with an individual's unique head and ears anatomy. The use of HRTFs compliant to the subjects anatomical traits is crucial to ensure a personalized and unique spatial experience. This work proposes the implementation of an HRTF individualization method based on anthropometric features automatically extracted from ear images using a Convolutional Neural Network (CNN).
\medskip

Firstly, a Convolution Neural Network is implemented and tested to assess the performance of machine learning on positioning landmarks on images of the ear. The I-BUG dataset, containing ear images with corresponding 55 landmarks, was used to train and test the neural network. Subsequently, 12 relevant landmarks were selected to correspond to 7 specific anthropometric measurements established by the HUTUBS database. These landmarks serve as a reference for distance computation in pixels in order to retrieve the anthropometric measurements from the ear images.
\medskip

A focus on the pinna mesh from 3D head meshes was conducted to retrieve 2D images of the ear, landmarks were annotated, distances were then computed and compared to the original measurements from the HUTUBS database in order to obtain the conversion factors. These were used to convert in centimetres the anthropometric measurements in pixels computed from the selected landmarks in the subject ear image.
\medskip

Once the 7 distances in centimetres are extracted from the ear image, a best match method using the 7 distances vector is implemented computing the Euclidean distance for each set in a database of 116 ears with their corresponding 7 anthropometric measurements provided by the HUTUBS database. The closest match of anthropometry can be identified and the corresponding set of HRTFs can be obtained for personnalized use. 
\medskip

The method is evaluated in its validity instead of the accuracy of the results. The conceptual scope of each stage has been verified and substantiated to function correctly. The various steps and the available elements in the process are reviewed and challenged to define a greater algorithm entity designed for the desired task. 
\medskip

Keywords : Spatial audio, HRTF individualization, Convolutional Neural Network, Anthropometric measurements, 3D head mesh, HUTUBS, CIPIC.


\newpage
\chapter*{Acknowledgements}

I would like to thank my promoter, Thierry DUTOIT for his enthusiasm and passions which sparked my interest in audio and acoustics. He has been an inspirational teacher and academic promoter throughout my university journey. 
\bigskip

I am thankful to my second promoter, Loïc REBOURSIERE for his constant support and guidance, as well as his advice, comments and opinions on the development of my thesis. His insights, advice, and constructive feedback have been invaluable.
\bigskip

I would also like to thank Sohaib LARABA and Prernna BHATNAGAR for their help and support in advising specific artificial intelligence technologies for my thesis.

\tableofcontents
\listoffigures
\listoftables
\makenomenclature
\printnomenclature
\newpage
\pagenumbering{arabic}

\chapter*{Introduction}

For a number of decades, Spatial audio and 3-Dimensional (3D) rendering have been present within our society. Extensive research and studies have been conducted which aim to provide the most realistic immersive audio experience for listeners. The goal of achieving true sound spatialization is a not new concept, contemporary advancements, particularly those utilizing artificial intelligence, have the potential to drive innovative progress.
\medskip

Video games, music, movies and virtual reality are at the core when exploring novel approaches aiming to enhance immersive experiences. Recently, the domain of immersive sound has expanded significantly enabling realistic 3D experience to be tailored for each individual. 
\medskip

The complexity of sound behaviour such as propagation and perception involves intricate acoustical principles and auditory cues for the human to perceive the direction of sound sources and feel immersed in a virtual environment. These complex auditory cues are all contained in one set of specific filters. 
\medskip

Head-Related Transfer Functions (HRTF) are unique acoustic filters which characterize how sound waves from various directions interact with the anatomy of the human head. This includes the effects of the shape, size, and position of the ear, as well as the influence of the head and torso. HRTFs play a crucial role in creating immersive spatial audio, enabling humans to perceive sound locations accurately in a 3-Dimensional space.
\medskip

It is possible to generalize these unique acoustic filters to create models which represent a broader range of human ear anatomies, leading to generalized HRTFs. This provides greater opportunities for more efficient and practical solutions with spatial audio applications, where personalized measurements might not be feasible on a large scale.
\medskip

While the idea of generalized HRTFs holds great potential, it is important to acknowledge that this approach might not be universally applicable. Individual auditory perception and anatomical features are unique and, therefore, a one-size-fits-all solution might not capture the nuances required for an optimal spatial audio experience for every listener. This leads to the consideration of the balance between the convenience of generalized models and the potential benefits of personalized HRTFs in supplying the full spectrum of human diversity in auditory perception.
\medskip

Achieving individualized HRTFs for each person presents challenges in terms of accessibility and feasibility. Traditional methods of HRTF measurement which involve extensive equipment and specialized procedures, may not be accessible to everyone. Moreover, the process can be time-consuming and labour-intensive. This is where it becomes interesting to explore automated approaches and by developing automated techniques for HRTF individualization, it has the potential to democratize access to high-quality spatial audio experiences. This could open doors for people all over the world to benefit from immersive auditory environments, regardless of their technical expertise or resources, thus bridging a gap in accessibility and ensuring an inclusive auditory landscape for everyone.
\medskip

Therefore, the primary objective of this work is to focus on a process to automate the HRTF individualization by designing the easiest way for someone to access their own unique set of functions. Related works and published papers will be reviewed, and a specific approach will be chosen and detailed.   To achieve this, artificial intelligence techniques will be integrated alongside Python scripting to establish an efficient computational pipeline which could potentially be adapted for universal applications. The ultimate focus is to offer users a simplified way to obtain their unique HRTF set, significantly improving the individualization process.
\medskip

This work will be dispatched in various sections. A state of the art will introduce spatial audio, rendering techniques and the concept of Head-Related Transfer Function (HRTFs) in Section 1. In Section 2, generalized HRTFs as well as their developments and their uses along with benchmark models will be discussed. The importance of individualized HRTFs and scientific approaches will be addressed in Section 3. The layout of the algorithm will be explained in Section 4 and further developments will be unveil in Section 5.

\setcounter{chapter}{0}

\chapter{State of the art}
\section{Introduction}
The 3-Dimensional audio immersion is a technology which enables a listener to be immersed in an environment using the hearing sense. Human beings are capable of processing sound waves into neuronal signals in the brain, which enables them to sense the pressure waves of sound surrounding them in their environment. 

\medskip

The human has two ears, one on each side of his head, these two ears allow physics phenomena to happen and enable him to build intrinsic computation mechanisms able to perceive which direction the sound is coming from. A human being is constantly exposed to an environment where sound and noise are present, from a bird song , a fly in close proximity or the breath sound of someone close by, to a door slamming, raised voices, a motorcar passing by, to the blairing of a train. 
\medskip

This complex mechanism is initiated at the early stage of childhood, when the infant discovers his environment, perceives sounds, and attempts to locate its source. That is how the baby can become aware of his surroundings. localization of sound is an intrinsic learning and begins as early as a few months old for infants, as they demonstrate the rudimentary ability to orient towards sounds and localise them within their environment. They further develop their innate mechanisms and sensory experiences which refine their sound localization skills. Through exposure to a variety of sounds and their spatial locations, children learn to associate auditory cues with specific directions and distances. They rely on active head movements to gather additional auditory information, such as changes in sound intensity and spectral cues. As their auditory system matures, the neural pathways responsible for sound processing become more specialised, contributing to the refinement of sound localization abilities. Significant advancements in sound localization skills typically occur during the first few years of life, with further improvements continuing throughout childhood and adolescence \cite{children}.
\medskip

Spatial audio aims to reproduce the human environment by immersing him in a digital world and tricking his hearing sense. The sound source is not at a real distance and direction from the human, it is produced and processed in order for the brain to perceive it this way. The goal of spatial audio is thus to reproduce a 3-Dimensional (3D) environment where the human will perceive a totally encompassing sound experience in all possible incoming directions and intensities.


\section{Spatial audio}

Spatial audio aims to replicate the way sound behaves in a real-world environment, enhancing immersion and realism in audio content. The concept of spatial audio may be applied to various fields, including music production and live staging, virtual reality (VR), augmented reality (AR), gaming, and cinema. It can enhance the sense of presence and the feeling of immersion in VR and AR experiences, creating a more realistic and engaging environment. In gaming, spatial audio can provide crucial positional audio cues, improving game play and situational awareness. In music production, it enables a more immersive and enveloping listening experiences with surround vocals and instruments.
\medskip

Spatial sound is an enhanced audio experience in which sounds can flow around, above, and below you in a 3-Dimensional virtual space which simulates a more realistic environment. Compared to the surround and stereo sound experiences, spatial sound tends to be an immersive atmosphere that traditional channel-based surround sound formats cannot replicate \cite{xbox}. The use of a specific number of speakers to reproduce audio in various directions is initially investigated with channel-based rendering. 
\medskip

\begin{figure}[!htbp]
     \centering
     \includegraphics[width=15cm]{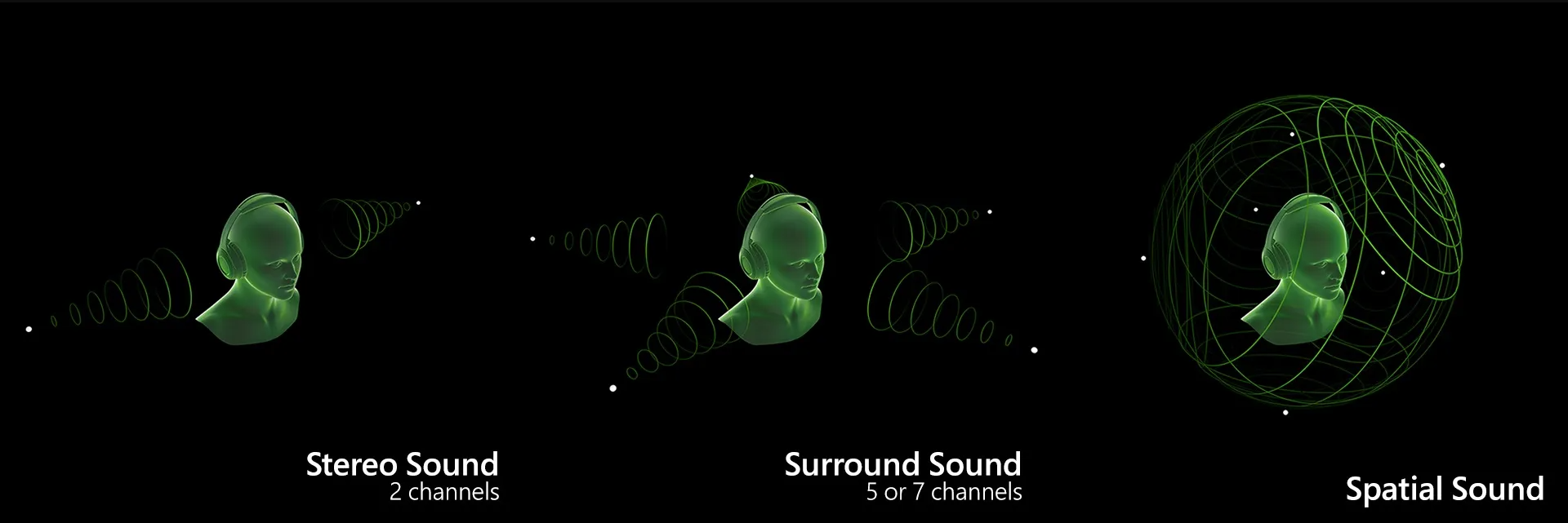}
     \caption{3D representation of spatial sound \cite{xbox}}
    \label{xbox}
\end{figure}

\subsection{Channel-based rendering}

Channel-based rendering is a sound spatialization technique used in audio environments where audio signals are processed and distributed to multiple channels to create the illusion of sound sources coming from specific directions and distances. Each channel corresponds to a speaker or an audio output, and by adjusting the amplitude and delay of the signals sent to each channel, the listener perceives sound sources as coming from different locations around him.
\medskip

The usual channel-based rendering techniques include stereo (2 channels), quadraphonic (4 channels), 5.1 surround sound (6 channels), 7.1 surround sound (8 channels). It is often limited to that specific number of channels as increasing the number of speakers becomes impractical in terms of setup complexity and budget. 
\medskip

With an increased number of source directions, spatialiazitaion of sound can bypasses the limitation of a fixed number of sound sources by utilizing specific techniques including binaural rendering, ambisonics and wave-field synthesis.

\subsection{Binaural rendering}
Binaural rendering aims to create a realistic and immersive 3D listening experience through dual-channel hardware such as headphones or earphones. The goal is to replicate the perception of sound as it would be heard in a 3-Dimensional space. Binaural rendering uses auditory cues to create the illusion of sound coming from specific directions and distances. It utilizes the two channels available from headphones and earphones, which are left and right signals. It creates the illusion of sound coming from different directions and distances. 
\medskip

The concept of binaural rendering considers the unique characteristics of each listener's head torso and ears with the use of Head-Related Transfer Functions (HRTFs) as well as other auditory cues such as interaural time differences (ITD) and interaural level differences (ILD) which result from the slight time delays and level disparities between the sounds reaching the individuals ear.
\medskip

\nomenclature{\(HRTF\)}{Head-Related Transfer Function} 
\nomenclature{\(ITD\)}{Interaural Time Difference} 
\nomenclature{\(ILD\)}{Interaural Level Difference} 

Binaural rendering is achieved by capturing or generating spatial audio content and then applying HRTFs by convolution to the audio signal to simulate the acoustic cues which would occur when sound reaches the listener's ears. The convolution of an audio signal and a filter involves applying the filter to the audio signal to modify its spectral characteristics. The filter would differ depending on the direction of the incoming sound. Binaural rendering simulates these cues to create the illusion of sound coming from specific directions and distances. This is achieved by processing the audio signals to include the appropriate time delays and level differences for each ear. When these processed signals are played back through headphones or earphones, our brains interpret the cues, creating a realistic and immersive 3D audio experience.
\medskip

An alternative approach to spatial audio rendering involves the use of virtual audio techniques, such as Ambisonics.


\subsection{Ambisonics}

Ambisonics is a technique for capturing, encoding, and reproducing immersive audio which aims to provide a 3-Dimensional sound experience in a complete spherical sound stage. It is based on the concept of soundfield representation, which considers not only the direction and position of sound sources but also their interaction with the surrounding environment. 'Ambisonics' is derived from the combination of the words 'ambi' meaning "all around" and "phonics" meaning 'sound'. Ambisonics is a technology for representing and reproducing the sound field at a given point" \cite{ambisonics1}.
\medskip

The sound field is represented by spherical harmonics, which are mathematical functions which describes the distribution of sound energy across various directions and frequencies. These spherical harmonics encode the intensity and directionality of sound sources in a 3D space, allowing for the accurate representation of sound localization and spatial cues. This enables for the capture and encoding of a complete sound scene, including both direct sound and spatial characteristics of the acoustic environment \cite{ambisonics2}.
\medskip

Ambisonics offers the advantage of scalability, accommodating various order levels in spherical harmonics. It spans from first-order, which utilises four channels, to higher-order Ambisonics (HOA) with additional channels. This scalability enables a finer spatial resolution, allowing for the capture of intricate sound scenes with enhanced spatial detail using higher-order Ambisonics. Encoding and decoding ambisonics spatial audio requires specialised algorithms. The quality of the spatial reproduction is affected by the number of channels used; higher-order ambisonics with more channels providing a greater spatial accuracy, but demanding more speakers and computational resources.

\medskip

The concept of Ambisonics can be converted to binaural audio for headphone playback, providing a personalized and spatialised sound experience but does not take into consideration the listener torso, head, and ears characteristics like the HRTF used in binaural rendering. 

\subsection{WFS : Wave-Field Synthesis }
\nomenclature{\(WFS\)}{Wave-Field Synthesis} 

Wave-Field Synthesis (WFS) is a spatial audio technique which strives to achieve an immersive auditory experience by accurately reproducing sound fields within a given physical space. Unlike conventional stereo or surround sound systems, WFS goes beyond channel-based rendering and binaural techniques. It operates on the principle of creating virtual sound sources throughout the listening area, simulating the behavior of sound waves in a natural environment. This is achieved by using an array of individually driven loudspeakers distributed in a two-dimensional plane, which are often surrounding the listener. By precisely controlling the phase, amplitude, and timing of the audio signals played through each loudspeaker, Wave-Field Synthesis generates complex interference patterns which mimics the way real sound waves interact and propagate. This results in a realistic 3D audio experience where listeners perceive sound coming from all directions, as if it were produced by actual physical sources in the environment. 
\medskip

The number of achievable source directions in spatial audio techniques like Wave-Field Synthesis is theoretically unlimited. However, the practical implementation of WFS might be constrained by factors such as the amount of available loudspeakers in the array, the spatial resolution of the system, and the computational complexity involved in accurately simulating a large number of sound sources. 

\section{Acoustic principles}

When hearing a sound, several acoustic principles come into play. These include sound wave propagation, reflection, diffraction, absorption, and resonance. These acoustic principles collectively contribute to the perception of sound. They shape the way sound waves travel, interact with the environment, and reach our ears.

To understand the complexity of how a sound is modified from its origin to reaching our eardrums, several acoustic principles have to be highlighted and explained. 

\subsection{Sound wave propagation}

In the atmosphere, sound waves are generated by a vibrating source and travel through air as a series of high-pressure regions called compressions and low-pressure regions known as rarefactions. It is a chain of acoustic pressures created by the movement of a surface from the smallest to the largest size. The rate at which these sound waves oscillate in pressure denote their frequency. The number of cycles of pressure oscillations in one second is referred to as the measurement unit known as a hertz (Hz). Humans have the ability to perceive sound frequencies spanning from 20 Hz (e.g. deep notes from a large pipe organ) to around 20 000 Hz (e.g. the chirping of a high-pitched bird). These sound waves propagate in different directions depending on the shape of the source which also define the shape of the wavefronts as the wave travels through the air away from its source. Wavefronts are the surfaces representing the continuous peaks of sound waves propagating through a medium indicating the points of equal phase and pressure. 
\medskip

The succession of compressions and rarefactions move outward from the source. The energy associated with each compression and rarefaction spreads out over an expanding spherical wavefront, signifying that the intensity decreases with distance because the same amount of energy is spread over a larger area. Sound waves can be reflected, refracted, diffracted, and absorbed as they encounter different materials and boundaries. 


\subsection{Acoustic reflection}

Reflection occurs when sound waves encounter a surface, a portion of the energy bounces back. Reflected sound waves can interact with other sound waves, leading to the formation of echoes and reverberation \cite{Acoustic1}. Reverberation consists of sound waves which reflect multiple times within an enclosed space, resulting in a persistence of sound even after the original sound source has stopped. The energy of multiple reflections significantly affects the persistence of sound as the energy decreases when reflecting off surfaces. The rate at which the energy decreases impacts the length of time the sound persists in an environment. When the surface is flat and smooth, the angle of reflection is the same as the incident angle which is the angle of the incoming sound wave with respect to the surface's normal. However, when the surface is not smooth and is irregular, the irregularities cause diffuse reflection and the sound waves are scattered in various directions. 
\medskip

The nature of the surface influences the amount of reflected acoustic pressure energy. Some surfaces tend to reflect sound more efficiently as they are hard and smooth like polished walls or glass. Porous or soft surfaces, for example foam and fabric absorb the acoustic pressure, resulting in reduced reflection. 
\medskip

Strong reflection (i.e. reflections happening on hard and smooth surfaces) can cause echoes and reverberation depending on the time delay between the direct sound and the reflected sound reaching the listener's ears.In environments where accurate sound reproduction is desired, controlling reflections through acoustic treatments or strategic placement of sound-absorbing materials can improve sound quality and clarity.

\subsection{Acoustic diffraction}

Diffraction refers to the bending or spreading of sound waves as they encounter an obstacle or pass through an opening. It allows the sound to reach areas which are not directly in the line of sight of the sound source, enabling us to hear sounds even when obstacles are present. Diffraction occurs because the variations in air pressure cannot go abruptly to zero after passing the edge of an object due to the compressions and rarefactions in the sound wave \cite{Acoustic1}. The specific folds and ridges of the ears cause sound waves to diffract and bend as they enter the ear canal from any direction.

\subsection{Acoustic absorption}

Absorption occurs when sound waves encounter a physical material which absorbs a partion of their energy. Different materials have varying degrees of sound absorption, and softer and more porous materials tend to absorb more sound energy. Absorption reduces sound reflections and helps control reverberation in a space \cite{Acoustic3}.

\subsection{Acoustic resonance}

Resonance is a phenomenon in which an object or medium vibrates at its natural frequency in response to an external sound source. When an object resonates, it amplifies certain frequencies which can affect the overall perception of sound. Resonance is often observed in musical instruments and architectural spaces. 
\medskip

In the middle ear cavity (see Section \ref{ear}), resonance plays a crucial role as the ossicles and the structures have specific resonant frequencies. This phenomena enables the amplification of certain frequencies of sound, thus enhancing our ability to detect and perceive different pitches \cite{Acoustic3}.

\section{Human sound localization cues}

Equipped with two ears on the opposite side of the head, Humans have the ability to perceive sound waves. This benefit enables different mechanisms to take place which contribute to the spatialization awareness of sound. These mechanisms are called localization cues and are intrinsic to humans as they are learnt during the early stages of life. 

\subsection{The ear canal} \label{ear}

When sound waves enter the human ear canal and reach the eardrum, the tympanic membrane vibrates in response. These vibrations are then transmitted to the middle ear where the malleus, the incus, and the stapes (commonly named the three small bones) ossicles and amplifies the sound vibrations \cite{Acoustic1}.

\begin{figure}[!htbp]
     \centering
     \includegraphics[width=10cm]{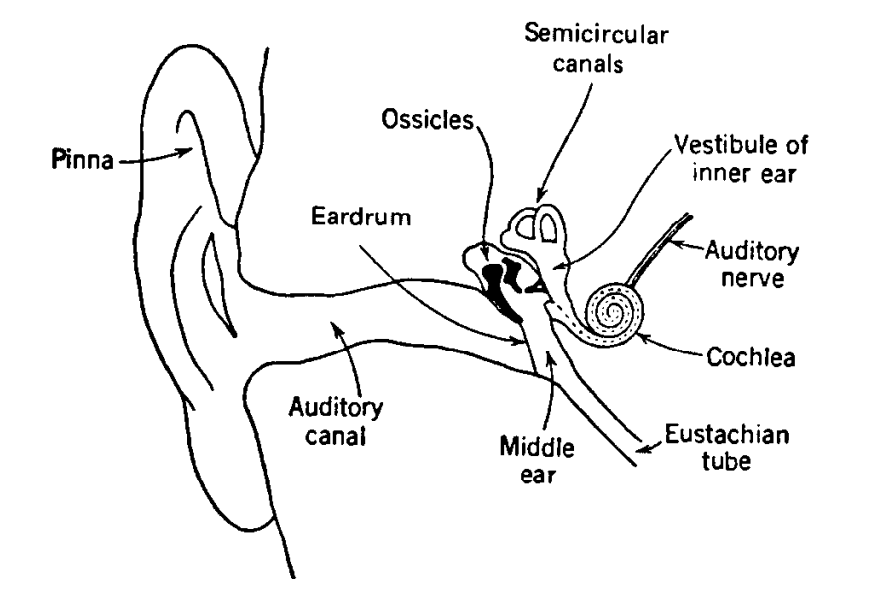}
     \caption{Sketch of the ear \cite{Acoustic1}}
    \label{AcouticEar}
\end{figure}

As the amplified sound vibrations reach the inner ear, they stimulate the hair cells in the cochlea, which convert mechanical vibrations into electrical signals which can be processed by the auditory system and interpreted by the brain.

\subsection{ILD : Interaural Level Difference}
"The ILD represents the intensity difference between the sounds received by the left and the right ears" \cite{HRTF}. Interaural level difference refers to the difference in sound level between the ears, it is the result of the sound source being located at a certain angle from the head causing the sound waves to arrive at each ear with different intensities. The brain compares the differences in sound level and uses it as information to determine the direction of the sound source \cite{HRTF}.
\medskip

The intensity of the sound which reaches the ears and is perceived as the sound level, provides cues on the distance of the source. 


\subsection{ITD : Interaural Time Difference}

"The ITD represents the time difference between the sounds
arriving in the left and right ears" \cite{HRTF}. Interaural Timing Difference refers to the difference of time arrival of a sound wave between the ears. It is caused by the sound source being located at a particular angle from the human head and the ears being located on opposite sides of it. ITD plays an important role as a sound localization cue particularly in low-frequency sounds in which wavelengths are larger and diffract around the head, resulting in minimal interaural level difference (ILD) \cite{HRTF}. "Interaural time difference (ITD) is the major cue for source azimuth (direction on the horizontal plane), and that ITD is probably encoded mostly by low-frequency auditory neurons" \cite{ITD1}.
\medskip

"There seems to be agreement that source elevation (direction above or below the horizontal plane) is cued by spectral peaks and notches, produced by pinna filtering , that occur mainly at frequencies above 5kHz" \cite{ITD1, ITD2}.

\medskip

These cues are based on the amplitude and time domain but the frequency domain also has an influence of how a human perceives a sound. A very important localization cue that takes into account the effect of the morphology of the human being including his torso, head and ears is the Head-Related Transfer Functions (HRTFs). 

\section{HRTF : Head-Related Transfer Functions} \label{HRTF}

The morphology of the human affects the way a sound wave propagates on its surface. The torso, head, ears, the ear cavity and the conduct to the eardrum act as filters to sound in all incoming directions. All the different incoming directions have a specific transfer function for each ear. Thus, it has the effect of a filter on the incoming signal which gives the human a spectral cue. The characteristics of sound which are related to the spread of energy across frequencies are known as spectral cues. These cues enable the human auditory system to identify the location of a sound and gain spatial information. Several graphics are shown in Figure \ref{hu_sub2} for visual representation of HRTF.  

\medskip

"While interaural time and level differences (ITD and ILD) are essential for distinguishing sources along the azimuth plane, spectral cues become more relevant when distinguishing source positions that vary in elevation" \cite{ITD3}. HRTFs represent the acoustic transfer function from the sound source position to the entrance of the ear canal of a human subject \cite{hrtfdef}. There is indeed a different HRTF for each ear and for all the possibilities of different angles of incoming sound. 
\medskip

HRTFs incorporate all the cues for sound localization, such as interaural time and level differences (ITD, ILD) and spectral cues which originate from scattering, diffraction, and reflection on the human pinna (the outer ear), head, and body \cite{hrtfcues}. HRTFs are typically measured under anechoic conditions, an environment with minimal or complete absence of sound reflection, at a sufficient distance and describe the complex frequency response as a function of the sound source position (i.e. azimuth and elevation). Imposing HRTFs onto a non-spatial audio signal and playing back the result using headphones allows for positioning virtual sound sources at arbitrary locations \cite{hrtfmagn}. 
\medskip

Humans have the ability to locate a sound source with two ears, and this remarkable binaural localization capability is largely attributed to the different filtering effects of the listener's head, pinna and torso on sounds originating from different directions in the frequency domain, which is transcripted by the HRTF. The HRTF dataset of one subject consists of the HRTF pairs of the left and right ears measured in various directions \cite{HRTF}. While ambisonics, binaural rendering  and WFS are spatial audio techniques offering generalized approaches for simulating the spatialization of sound, HRTFs are personalized acoustic filters aiming to reproduce the effect on sound of an individual head anatomy.

\subsubsection{HRTF history}

Researchers in the $mid-20^{th}$ century began investigating how the human auditory system localises sound sources in a 3D space. George Williams and James Beament, two Harvard researchers, conducted a pioneering experiment in the 1940s and 1950s to study sound localization cues and the concept of HRTF. In the 1970s and 1980s, HRTF research took a significant momentum when researchers Edgar Villchur and James M.Johnson \cite{audiobook} explored the role of the pinna, the external part of the ears, in sound localization. They realised that the shape and structure of the pinna were critical to capture the spectral filtering characteristics of sound as it reaches the eardrums from different directions \cite{audiobook2}. 
\medskip

Bill Gardner and Keith Martin published in 1994 their landmark work providing a comprehensive set of HRTF measurements which includes elevation, azimuth and distance information known as the MIT KEMAR HRTF dataset \cite{gardner}. In the late 1990s and early 2000s, with progress in digital signal processing and computational capabilities, researchers and companies began exploring personalized HRTFs. These individualized HRTFs aimed to create more accurate and immersive 3D audio experiences by capturing the unique acoustic characteristics of each listener's ears. The acquisition of these HRTFs were mainly realised with two techniques, measurements and simulation. 

\subsection{HRTF acquisition}

HRTFs capture the filtering effects of the torso, head and ears on sound as it reaches the eardrums from different directions in spaces. HRTFs have been researched and captured for half a century, and two main techniques have been used to capture sets of HRTFs. 
\medskip

Head-Related Transfer Functions are typically measured or simulated for a set of discrete directions in order to cover 360 degrees of azimuth angles (i.e. horizontal axis) and a range of elevation angles (i.e. vertical axis) \cite{sonicom}.

\subsubsection{HRTF measurements} \label{measure}

The measurement of these functions is performed using specialised equipment and techniques which involve the following steps. The subject is seated on a chair ideally in an anechoic chamber, and specific microphones known as in-ear microphones are placed in his ear canals in order to capture the sound pressure at the eardrums \cite{hrtfmagn}. A calibrated sound source using a speaker is positioned in various locations around the subject to represent the different azimuths and elevations angles. The sound sources then emit a signal which is often short broadband impulses or frequency sweeps, as the microphone records the impulses. A broadband impulse has a wide range of frequencies which cover a broad band of the frequency spectrum. A frequency sweep is a continuous waveform which steadily increases or decreases in frequency over a specified range within a specific duration. The recorded impulse responses are called Human-Related Impulse Responses (HRIRs), and they represent the acoustic response of the human head and ears to an acoustic impulse in the time domain. The sound source is moved systematically to cover the desired range of directions around the subject, resulting in a measurement grid of HRIRs at discrete spatial locations \cite{children, personalization}.
\medskip

\begin{figure}[!htbp]
     \centering
     \includegraphics[width=8cm]{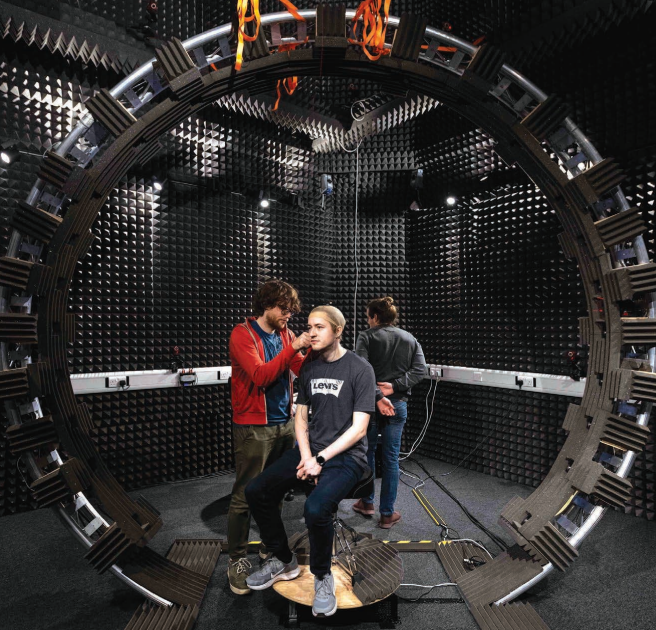}
     \caption{The HRTF measurement setup at Imperial College London, U.K. \cite{sonicom}. The subject is seated on a rotating chair equipped with ear-inserted microphones, encircled by an array of specifically positioned speakers emitting impulse responses.}
    \label{hrtf setup}
\end{figure}

\nomenclature{\(HRIRs\)}{Human-Related Impulse Responses}
\nomenclature{\(ADC\)}{Analog to Digital Converter}

The analogue signal (sound wave emitted by the speaker) is captured by the microphone and converted to a digital format using an analogue-to-digital converter (ADC). This process involves sampling the amplitude of the analogue signal at regular intervals and representing each sample as a digital value. During the analogue to digital conversion process, the continuous amplitude values of the analogue signal are converted into discrete digital values. This quantisation process determines the bit-depth of a .wav file, which defines the resolution of the digital representation. Common bit-depths include 8-bits, 16-bits, 24-bits, and 32-bits values. These values define the dynamic range of the audio file. The .wav file includes metadata and a header that contains essential information about the audio such as sample rate, number of channels, bit-depth, and audio length. HRIRs are stored in .wav files as they are a time-domain representation of the acoustic response of the human head and ears to an acoustic impulse from various spatial directions. HRTF is the frequency domain representation of HRIR and is computed by performing a Fourier transform on the HRIR data.  A Fourier transform converts a signal from the time domain to the frequency domain, resulting in a frequency spectrum which contains both magnitude and phase information for each frequency. The magnitude represents the amplitude, and the phase represents the phase shift experienced by that frequency component as it passes through the listener's head and ears \cite{survey, highres}.
\medskip

By applying the Fourier transform to each HRIR signal, it is possible to represent how the sound energy is distributed across different frequencies for different spatial directions and acquire the corresponding HRTFs.
\medskip

Other steps like calibration and post-processing are often required in order to have accurate measurements \cite{children}. Calibration involves adjusting the factors and parameters of the impulse response, amplifier gains, room reflection, microphone responses and all the factors that can influence the accuracy of the measurement. Post-processing of the measured HRTFs is used to remove any unwanted artefacts or noise, as well as to interpolate the data of different directions for smooth variation between measured locations \cite{technical}.

\subsubsection{HRTF simulation} 

Head-Related Transfer Functions can also be acquired by simulation which involves the following steps. An anthropometric model is chosen, it can either be a computed average human head and ears shapes or a specific human head representing an individual morphology. The anthropometric term refers to the measurement and analysis of human body dimensions and physical characteristics. It is imperative that both of these models provide precise geometric information about the head and ear morphology and structure to be represented correctly in simulation. Numerical methods such as boundary element methods (BEM) \cite{Katz} for the harmonic domain or finite difference time domain (FDTD) \cite{Ziegelwanger} for the time domain are used to simulate sound propagation around the subject's head and ears, while considering the acoustic principles of sound propagation and the acoustic properties of the head and ear \cite{survey, pinnainfluence}. 
\medskip

\begin{figure}[!htbp]
     \centering
     \includegraphics[width=6cm]{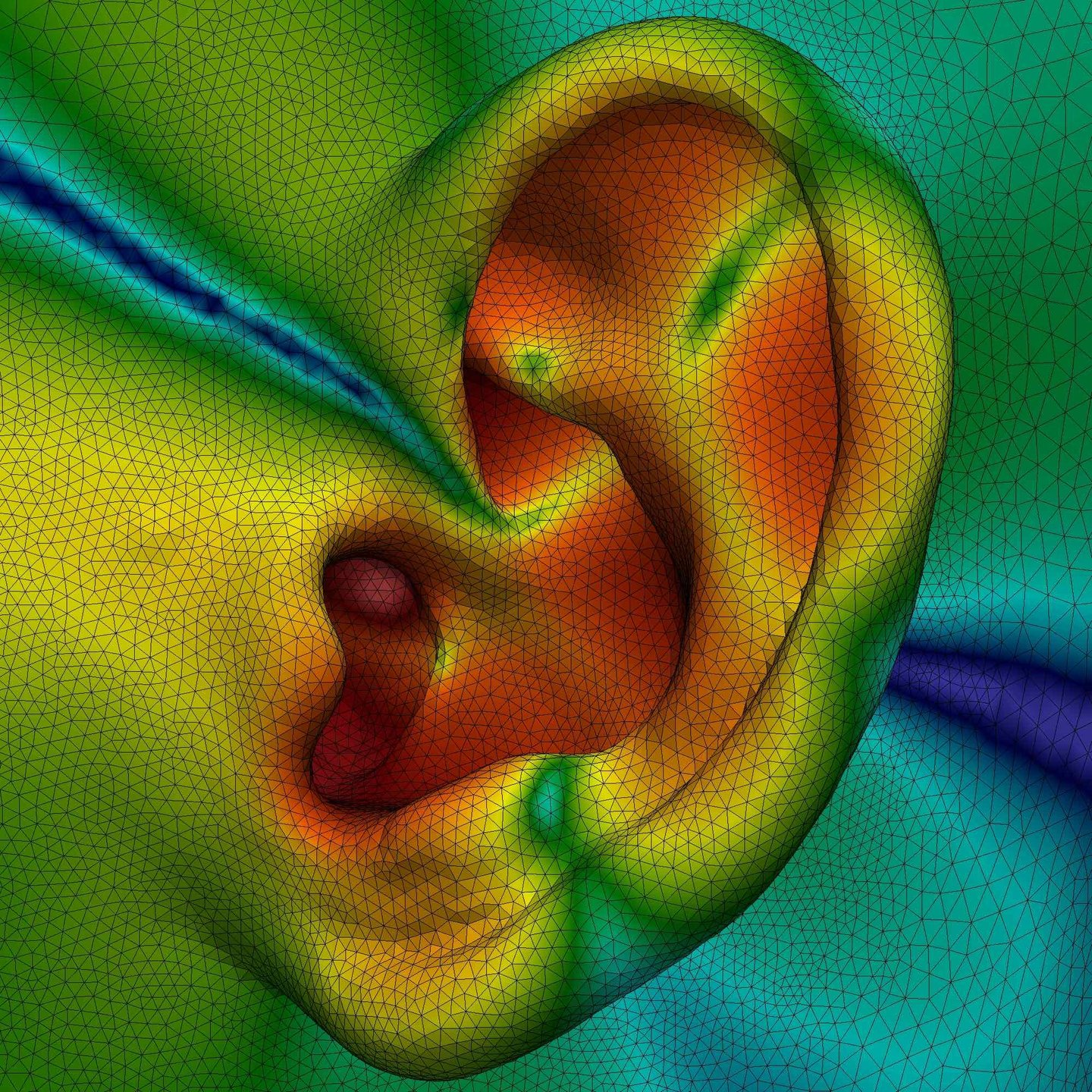}
     \caption{Ear pinna mesh represented in Mesh2HRTF \cite{mesh2hrtf}; the colouring represents the sound pressure level}
    \label{Mesh2HRTF}
\end{figure}

A 3D model of the listener's head and ears is created from the anthropometric data of the subject and represents the shape and geometry of the subject's head, outer ear, and ear canal. The 3D model is discretised into a mesh of smaller elements, such as triangles or quadrilaterals. BEM and FDTD then formulate acoustic wave equations as integral equations on the surface of the mesh elements describing how the sound waves interact with the head and ear surfaces \cite{highres}.
\medskip

The head and ear shape and tissue impedance (i.e. resistance of the tissues to sound waves), the speed of sound, and the air density all have an influence on how sound waves are reflected, refracted and diffracted. The HRTF is computed from the modification between the original source signal and the signal computed at the ear canal of the subject. It is defined as the filter applied to the original signal to obtain the signal computed at the subjects ear. Similarly to HRTF measurements, simulations are performed for a specific range of directions around the virtual model, covering azimuth and elevation angles, resulting in a spatially sampled HRTF dataset. 
\medskip

\begin{figure}[!htbp]
     \centering
     \includegraphics[width=8cm]{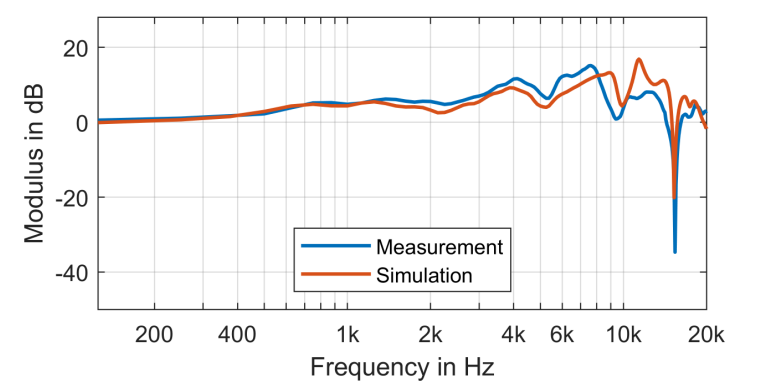}
     \caption{Comparison between measured and simulated HRTF of a subject left ear at $\phi = 150^{\circ}$ in the horizontal plane \cite{Braren}}
    \label{comp}
\end{figure}

The simulated HRTFs are further processed to compensate any resonances, smooth the responses, and optimise the spectral and spatial accuracy of the simulation. These post-processing adjustments aim for smoother results and better accuracy of the HRTF results. A validation between the simulated HRTFs and measured HRTFs is possible to ensure the accuracy and reliability evaluating any discrepancies and irregularities \cite{cross-eva, Braren}, as shown in Figure \ref{comp}.

Mesh2HRTF \cite{mesh2hrtf} is a software tool which allows to load a 3D head mesh (i.e. a 3D model) and simulate how sound waves interact with the individual's specific anatomy. Mesh2HRTF enables the creation of personalized Head-Related Transfer Functions (HRTFs) for individuals from high resolution 3D scans. The disadvantage and limitations of Mesh2HRTF is its complexity and resource-intensive nature. Additionally, the accuracy of the personalized HRTFs generated by Mesh2HRTF relies heavily on the quality of the 3D scans which can be challenging to obtain without equipment. An example of an ear mesh in Mesh2HRTF in represented in Figure \ref{Mesh2HRTF}.

\nomenclature{\(BEM\)}{Boudary Element Methods} 
\nomenclature{\(FDTD\)}{Finite-Difference Time-Domain} 

\subsection{HRTF formats}

Once the HRTFs are acquired, they are stored in specific formats in order to be used with ease and efficient processing. Various formats exist for storing HRTFs data and datasets. The format chosen depends on the application of the HRTFs required by the user. 
\medskip

Storing HRTFs in an efficient, accurate, detailed, and organised manner enable accurate and realistic sound spatialization in spatial audio applications. One distinctive format in the HRTF domain is the SOFA format.
\medskip

\subsubsection{SOFA : Spatially Oriented Format for Acoustics}
\nomenclature{\(SOFA\)}{Spatially Oriented Format for Acoustics} 

HRTFs have been measured for many years by several laboratories and are usually stored in each laboratory specific formats. These native formats have benefits for each laboratory but are not designed for data exchange purposes by the fact that they are not compatible with each other. 
\medskip

The need to represent and store HRTF data in a consistent and versatile format led to the development of a standardised format called SOFA, a standard of the Audio Engineering Society (AES) \cite{AES}. The spatially orientated format for acoustic storage and exchange are not only HRTF datasets, but also more complex spatial audio data and metadata. SOFA defines conventions which specify the unique structure and organization of the data ensuring dataset formats for integration into audio processing software and research tools. 
\medskip

\nomenclature{\(AES\)}{Audio Engineering Society} 

The metadata represent information such as the subject's anthropometric measurements, measurement setup details (e.g. loudspeaker positions, subject and loudspeakers positions), the sampling frequency, and various calibration parameters. Units of measurement are also specified to ensure consistency across datasets and enable conversion between different units. This information allows for spatial rendering and thus sound localization.
\medskip

\nomenclature{\(XML\)}{Extensible Markup Language}

The structured data of SOFA files are based on the Extensible Markup Language (XML), which allows for organization of metadata and measurement data and are readable by both humans and computers. Data arrays represent in both the time and the frequency domains impulse responses, Head-Related Transfer Functions, or any other spatial audio data. The measurements are represented in a 3D Cartesian coordinate system with elevation, azimuth, and frequency as dimensions. 
\medskip

\nomenclature{\(API\)}{Application Programming Interface}

SOFA has its own Application Programming Interface (API)\footnote{Sofaconventions.org : Software and APIs. Consulted in February 2023. \url{https://www.sofaconventions.org/mediawiki/index.php/Software_and_APIs}}, "pysofaconventions" stands for Python SOFA Conventions which facilitate reading and manipulating SOFA data in Python coding language.
\medskip

The standardised nature of SOFA files has made this format to be widely adopted in spatial audio research, HRTF database creation, and virtual auditory display applications. Several HRTF datasets using the SOFA file format are available on the sofa convention website\footnote{Sofaconventions.org : Files. Consulted in February 2023. \url{https://www.sofaconventions.org/mediawiki/index.php/Files}}.

\subsubsection{MATLAB}

HRTF measurements can be stored in MATLAB data files that contain the impulse responses for different directions and frequencies. MATLAB is known as a high-level programming language and interactive environment designed for numerical computation, data analysis, signal processing, and visualisation. Thus, it allows researchers and engineers to process, analyze and manipulate HRTF data. 
\medskip

MATLAB offers the possibility to realize modifications on HRTF such as filtering, resampling, and spatial interpolation in order to prepare HRTFs for binaural rendering or individualized audio rendering. The visualization tools allow easy-understanding representation and enables visualization of HRTF measurements, compare datasets and acquire clear and informative presentation \footnote{MathWorks.com : Interpolate HRTF. Consulted in February 2023. \url{https://nl.mathworks.com/help/audio/ref/interpolatehrtf.html}}.
\medskip

Before going any further in the HRTFs scientific aspect, applications and technologies based around spatial sound and HRTFs are presented in the next two sections. These examples help to understand why such specific acoustic filters are so important in creating realistic and immersive 3D sound experiences.

\section{Application of HRTF}

Head-Related Transfer Functions (HRTFs) have become a key element in the area of spatial audio, transforming the way humans sense and experience sound in a variety of applications. The ability to simulate sound localization and create a sense of 3-Dimensional auditory space has led to diverse applications of HRTFs in areas such as music, movies, gaming, virtual reality, augmented reality, hearing aid devices and more.

\subsection{Music industry}
Audio in the music industry uses spatialization of sound to elevate the listener experience. This means the positioning of different instruments, vocals, and drums all around the listener at varying degrees and intensities \footnote{AppleInsider.com : Why Spatial Audio is the future of the music industry, even if you hate it by Tyler Hayes, January 29 2023. Consulted in February 2023. \url{https://appleinsider.com/articles/23/01/29/why-spatial-audio-is-the-future-of-the-music-industry-even-if-you-hate-it}}. Several artists such as Billie Ellish, Sam Smith and more produce music in studio for spatial audio and conceived for the immersive listening experience of their production. 
\medskip

The acoustic sound of a concert could be reproduced via spatial audio and the use of HRTFs, the listener could navigate into a concert environment and hear all around him with the artists performing in front of him, the crowd cheering by his side and the resonance of a concert venue behind him. The possibility of producing music in a virtual environment enables numerous possibilities of spatial experiences for the listener. HRTFs take the stereo audio and turn it into an immersive experience of sound.

\medskip

Spatial audio can be achieved with the use of headphones and HRTF functions correctly applied to the sound generating from every direction. The immersive audio experience of listening to music could then be available to anyone who owns a pair of headphones, a set of HRTF and the compatible software and hardware. 

\subsection{Cinematic immersion}

Spatialization of sound in a movie theatre or in a home cinema setup has existed for a long time and has evolved extensively over the years. Several standards of immersive audio like Dolby Atmos, THX\footnote{THX is a quality certification and audio-visual technology company founded by George Lucas in 1983, originally created to ensure that movie theaters were capable of delivering high-quality sound and picture. Consulted in February 2023.\url{https://www.thx.com/}} and DTS\footnote{DTS : Digital Theater Systems is a similar brand providing high-quality immersive audio experience using encoding and decoding algorithms to create multi-channel surround sound. Consulted in February 2023. \url{https://dts.com/}} are aiming to provide spatial auditory experience with surround sound and atmospherics effects with the use of speakers placed all around the auditorium. 
\medskip

\nomenclature{\(DTS\)}{Digital Theatre Systems}

"The use of HRTFs and spatial audio techniques by filmmakers create immersive 3D surround soundscapes that enhance emotional impact, provide aural depth, and foster realism in movie scenes"\footnote{Netflix.com : Bringing New Features to Netflix’s Premium Plan by Rishu Arora, February 01 2023. Consulted in February 2023. \url{https://about.netflix.com/en/news/bringing-new-features-to-netflixs-premium-plan}}. HRTFs enable spatial positioning of audio sources around the audience, and thus add a sense of presence, heightening engagement in action sequences, and contributing to storytelling by guiding the viewer's attention.
\medskip

Spatial audio movies could be experienced at home with the use of headphones, a set of HRTFs, and the appropriate software. A greater audience could access a detailed 3D surround sound experience without using a home cinema speaker system.

\subsection{Gaming experience}

HRTFs are used in game audio engines such as Unity\footnote{Unity.com. Consulted in February 2023. \url{https://unity.com/}} or Unreal\footnote{Unrealengine.com. Consulted in February 2023. \url{https://www.unrealengine.com/en-US/}} to provide sound localization for virtual objects and characters. Players can identify the direction of the sounds in games for example footsteps, gunshots, or dialogue which enhances situational awareness and game-play immersion.

\subsection{VR : Virtual Reality }
\nomenclature{\(VR\)}{Virtual Reality}

Virtual reality is the term used for the experience of a digital representation of reality. VR is the term employed to describe any digital 3D representation that tends to mimic the reality. Virtual reality is accessible via specific headsets, which contain screens for the eyes and electronics to sense the user head movement in order to adapt the virtual world vision. 
\medskip

In this virtual reality, the sense of sight is the primary focus but a second sense can be considered for a more immersive and natural experience. The sense of hearing with the accurate position of virtual sound in a 3D space can be achieved with the use of HRTFs \cite{usability}. By applying a set of HRTFs on the audio signal from the virtual scene, virtual experiences of auditory immersion takes the user to a deeper level in the awareness of the 3D environment surrounding him. An HRTF set provides the magnitude, temporal, and spectral cues that the auditory system utilises for spatial perception \cite{predict}.
\medskip

\subsection{AR : Augmented Reality }
\nomenclature{\(AR\)}{Augmented Reality}

The possibility of increasing the information available in our reality and interacting with it is the purpose of augmented reality \cite{AR}. AR headsets can place information in the human field of view which in turn blends into our reality. If new information comes with sound, the use of HRTFs on the audio signal enables the user to perceive sound around him for a better immersive experience. 

\subsection{Hearing aid}

The sound processing in hearing aids and cochlear implants could be personalized by considering the shape of the person's head and ears with HRTFs for improved spatial perception of sound, leading to better speech intelligibility and sound localization. 

\section{Technologies using HRTF}

There are several standards and companies which contribute to spatial audio using HRTFs. These offer immersive sound experiences experimenting with different technologies, devices and algorithms to achieve excellence in immersive spatialization. Unfortunately, very little technical information is available on the way HRTFs are used. HRTF terms are not always brought forward and technical papers are rarely made public.

\subsection{Dolby Atmos}
Dolby Atmos, developed by Dolby Laboratories, introduces an object-based approach to audio technology which treats sound as individual objects. These objects can be precisely positioned and moved in a virtual 3-Dimensional space, transforming sound capture, mixing, and reproduction across entertainment mediums like music, movies, TV shows, and gaming. Sound objects are no longer confined to specific channels or speakers; they are freely placed around the listener, offering accurate localization and an immersive audio experience. Dolby Atmos employs HRTFs to position sound objects in a virtual 3D sound space, whether in a cinema, home theater, or mobile setup, ensuring seamless movement and an engaging auditory experience. This technology has gained popularity in cinemas, home setups, soundbars, and headphones \cite{Dolby2}.

\subsection{Spatial audio by Apple}

Apple's "Spatial Audio" creates a 3-Dimensional auditory experience for compatible devices like AirPods Pro and AirPods Max. It seeks to replicate the immersive ambiance of theaters and live performances by delivering a dynamically spatial listening experience \footnote{Apple.com. Consulted in February 2023. \url{https://www.apple.com/befr/}}. This is achieved through advanced algorithms and motion sensors which track the user's head and device position. By combining this data with directional audio filters and generalized HRTFs \footnote{Generalized HRTFs are HRTFs but based on an average human head and ears, rather than being specific to an individual's unique anatomy \cite{2}}, it simulates sound originating from specific directions in the surrounding space. While technical specifics of Apple's HRTFs aren't disclosed, their technology capitalizes on the H1/H2 and W1 chip synchronization, ensuring low-latency audio processing and support for multi-channel audio formats like Dolby Atmos.

\subsection{360 Reality Audio by Sony}

Sony's "360 Reality Audio" aspires to provide an immersive and lifelike music listening encounter. It crafts a 3-Dimensional auditory environment by situating individual instruments and vocals within a spherical space around the listener, using object-based spatial audio. To achieve this, Sony recommends compatible headphones or speakers which are strategically designed for 360 Reality Audio. These devices incorporate multiple drivers or speakers to deliver precise spatial sound reproduction. Furthermore, they consider the listener's unique  transfer function (HRTF) data, tailoring audio playback to create a personalized listening experience which aligns with the listener's sound perception \footnote{Sony.com : 360 Reality Audio. Consulted in February 2023. \url{https://electronics.sony.com/360-reality-audio}}.

\subsection{Unity and Unreal video game engines}

Video game engines are software platforms that provide tools and functionalities for creating and developing video games. They streamline the game development process by offering features such as graphics rendering, physics simulations, audio management, scripting, and more. Developers use game engines to build games for various platforms efficiently and effectively. Unity and Unreal are two video game engines which handle and provide spatialization of sound for gaming experience.
\medskip

Unity provides support for spatial audio through its "Spatializer" plugins. One of the plugins that Unity supports is the Google Resonance Audio Spatializer\footnote{Resonance audio.com. Consulted in April 2023. \url{https://resonance-audio.github.io/resonance-audio/}}, previously known as Google VR Audio Spatializer. Resonance Audio is designed to create an immersive 3D audio experience that includes HRTF processing to simulate sound patialization.
\medskip

Unreal Engine's audio system allows developers to use HRTFs to position sound sources accurately in 3D space. The engine supports binaural rendering, enabling realistic audio cues for players as they move within the game world. Unreal Engine also provides functionality for audio occlusion and obstruction which enhances the realism of the sound environment by considering objects and geometry between the listener and the sound source\footnote{Unreal.com. Consulted in April 2023. \url{https://www.unrealengine.com/en-US/}}.
\medskip

Both Unity and Unreal Engine aim to provide developers with the tools to create immersive and realistic audio experiences for games and interactive applications. By integrating HRTFs and spatial audio techniques, developers can enhance the auditory dimension of their projects, increasing player engagement and immersion in the virtual worlds which they create.
\medskip

Now that these applications and technologies have been introduced, let's delve into the broader scientific aspects of the Head-Related Transfer Functions. 

\chapter{Generalized HRTFs}

As explained in Section \ref{HRTF}, HRTFs are acoustic filters representing how sound interacts with an individual's head and ears. They can be generalized to provide standardized spatial audio experiences across a diverse group of listeners without the need for individual measurements, therefore called "Generalized HRTFs".
\medskip

HRTFs can be generalized so that the functions can be used for a large number of people based on common anthropometric measurements. Generalized HRTFs refer to a set of spatial audio filters which represent the average or standardized acoustic characteristics of the human head and ears. These generalized functions are computed or measured from an average morphology of humans. They are derived from measurements taken from a diverse group of individuals in order to capture common trends and characteristics found in human head and ear shapes.
\medskip

\begin{figure}[!htbp]
     \centering
     \includegraphics[width=10cm]{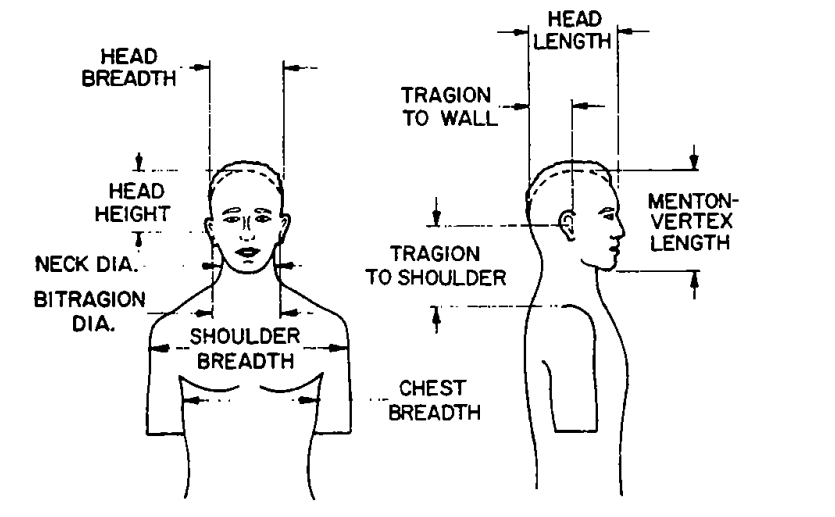}
     \caption{Anthropometric measures used in design of KEMAR \cite{Manikin}}
    \label{KEMAR1}
\end{figure}

Non-individual binaural synthesis engines often use the same generic HRTFs for all users and enable binaural rendering to provide accessible spatial audio immersion. HRTFs measured with the use of manikins such as the examples explained below provide a reference point for simulating the acoustic cues experienced by an average listener. They provide a more accessible solution by offering standardized templates, as shown in Figure \ref{KEMAR1} which can be used for a broader range of users without personalized measurements \cite{2}. Since they capture common acoustic characteristics found in human head and ear shapes, generalized HRTFs lead to a more standardized sound localization experience for multiple listeners. In the following sections, three different models used as reference anthropometry for the human will be presented and illustrated. These three manikins are especially valuable for spatial audio, virtual reality, and gaming, where accurate sound localization and an immersive auditory experience are essential.

\section{KEMAR} \label{HAT}
\nomenclature{\(KEMAR\)}{Knowles Electronics Manikin for Acoustic Research}
\nomenclature{\(MIT\)}{Massachusetts Institute of Technology}
\nomenclature{\(HAT\)}{Head And Torso}
\nomenclature{\(HATS\)}{Head And Torso Simulator}

The KEMAR (Knowles Electronics Manikin for Acoustic Research) is a human-sized artificial head and torso designed to represent the average human anthropometry. It is used for acoustic research and particularly for capturing transfer functions. It was designed at the MIT ( Massachusetts Institute of Technology) in 1972 and has become an accepted industry standard in the field of spatial audio, hearing-aid manufacturers, and research audiologists \cite{GRAS}.
\medskip

Crafted to mimic the average dimensions of the human head and ears, each aspect of the KEMAR manikin was carefully selected to embody those of a typical adult. Reputed as a standardized point of reference in spatial audio exploration, this manikin has become a foundation in research activities. HRTFs measured from the KEMAR manikin serve as fundamentals for creating generalized or standardized HRTF databases, which are crucial for achieving accurate sound spatialization across various applications.\cite{Manikin}. 
\medskip

\nomenclature{\(G.R.A.S.\)}{Grande Ronde Acoustic System}

\begin{figure}[!htbp]
     \centering
     \includegraphics[width=5cm]{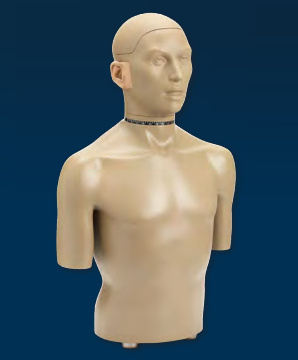}
     \caption{G.R.A.S. 45BB KEMAR Head $\&$ Torso \cite{KemarPoster}}
    \label{KEMAR2}
\end{figure}

On Figure \ref{KEMAR2}, the G.R.A.S. (Grande Ronde Acoustic System) inspired by the KEMAR manikin pattern.

It is one of the only manikins which offers the ability to adjust the ear-to-shoulder ratio, making it suitable for simulating both male and female average values. The manikin is equipped with microphones embedded in the ear canals which capture sound signals when they reach the eardrums. The built-in microphone is the 40AH Low-noise Ext. Polarized Pressure Microphone\footnote{Grasacoustics.com : GRAS 43BB Low-noise Ear Simulator System. Consulted in April 2023. \url{https://www.grasacoustics.com/products/ear-simulator-kit/traditional-power-supply-lemo/product/722-43bb}}. During the acoustic measurements using the G.R.A.S. 45BB KEMAR, the manikin is placed in the center of an array of speakers or a mobile speaker, then various impulse response sources are played around the manikin while the sound is recorded as it reaches the microphones within the ears. The manikin remains stationary as the source is mobile. It is therefore the same procedure detailed in Section \ref{measure}. The KEMAR generalized HRTFs can be found in sofa format\footnote{Sofacoustics.org : KEMAR HRTF. Consulted in April 2023. \url{https://sofacoustics.org/data/database/mit/}}. Another manikin which serves as a reference in the generalization of HRTFs is the Neumann KU100.

\section{Neumann KU100}
The Neumann KU100, also known as the "Binaural Head," is a human-sized manikin renowned for its replication of the average human head and ears. It was developed in the early 1970s by Georg Neumann GmbH, a pioneering German audio equipment manufacturer. The Neumann KU100 was designed for binaural recording purposes, aiming to replicate human hearing characteristics and capture spatial audio cues. The ears were designed to mimic the size, shape, and acoustic properties of human ears in an average anatomy approach. Equipped with in-ear binaural stereo microphones, it captures sound from 20 Hz to 20 000 Hz to cover the entire human auditory spectrum. The Neumann KU100 is a known tool for creating immersive auditory experiences and has contributed significantly to binaural audio research \cite{Neumann}.
\medskip

\begin{figure}[!htbp]
     \centering
     \includegraphics[width=6cm]{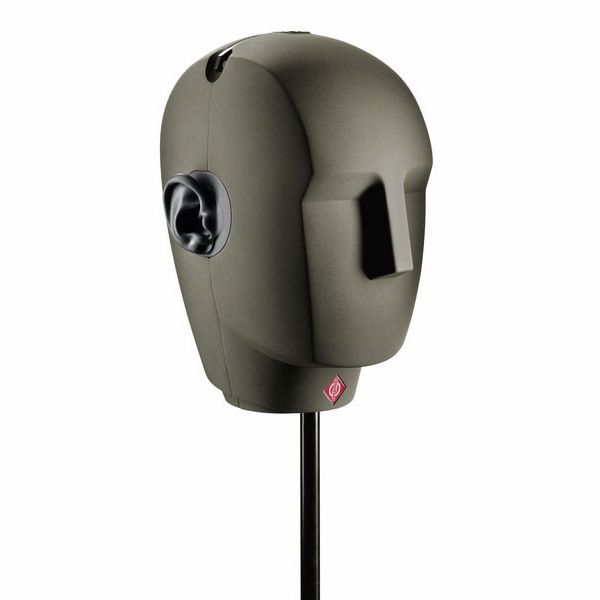}
     \caption{Neumann KU100 manikin \cite{Neumannjpg}}
    \label{Neumannjpg}
\end{figure}

While the KU100 captures binaural audio based on its artificial ears, the recorded HRTFs serve as a standardized representation of how sound interacts with a general human head and ear shape. The HRIRs of the Neumann KU100 are available online in SOFA and Matlab fromats\footnote{ Zenodo.org : Spherical Far-Field HRIR Compilation of the Neumann KU100 2020. Consulted in June 2023. \url{https://zenodo.org/record/3928297}}.

\section{Brüel \& Kjær Type 4128-C HAT} 

The Brüel \& Kjær Type 4128-C Head and Torso Simulator (HATS) ,introduced in the early 2000s, is an advanced anthropomorphic manikin (more recent)  used in audio and acoustic applications. Designed by Brüel \& Kjær, a Danish company specializing in sound and vibration measurement solutions, the Type 4128-C HATS aim to  offer accuracy and realism in the capture and reproduction of spatialized sound \cite{TYPE4128-C} similar to the two previous manikins. With its human-sized head and torso design, the Type 4128-C HATS provides a valuable tool for studying sound perception, evaluating audio devices, and optimizing audio system performances. 
\medskip

\begin{figure}[!htbp]
     \centering
     \includegraphics[width=7cm]{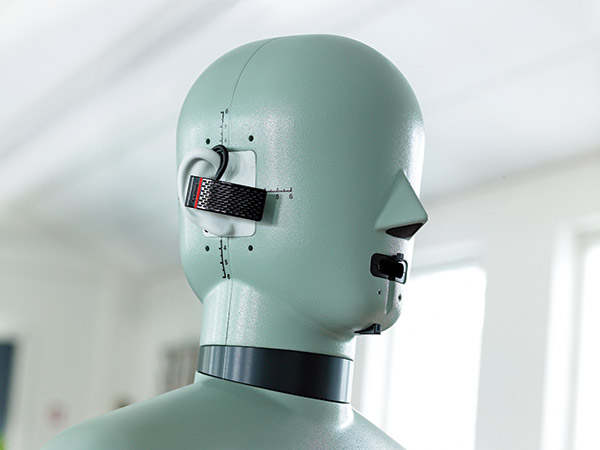}
     \caption{TYPE 4128-C Head And Torso Simulator (HATS) \cite{TYPE4128-C}}
    \label{TYPE4128-C}
\end{figure}

These manikins serve as references for capturing generalized HRTFs for a wide diversity of people. Other manikins and models of average human morphology exist and are used by companies seeking to obtain generalized HRTFs suitable for as many people as possible. While the pursuit of a common anatomical representation is important, individual morphology is also crucial as it defines the unique shape of a person's anatomy and its impact on sound. Although manikins with average head shapes provide valuable insight, they may not accurately represent the individual listener's unique ear shapes, sizes, and anthropometric variations which impacts their personal HRTFs. By following each person's unique anatomy, individual HRTFs can be considered and introduced. 

\chapter{Individualized HRTFs}

As mentioned earlier, Head-Related Transfer Functions (HRTFs) are essential for shaping our auditory perception and enabling precise sound localization in virtual 3D space. However, when using generalized HRTFs, several inadequacies may arise due to the unique anatomical traits of each individual. These consequences include front-back sound confusion, lack of externalization of sound, and sound localization accuracy degradation (particularly in elevation perception and to a lesser extent in horizontal perception). These consequences are explained in \cite{2, need, 4, 5} and mentioned in \cite{extracted}. To address these limitations, individualized HRTFs are becoming increasingly recognized as essential for personal spatial audio experience compared to using generalized HRTFs.
\medskip

As highlighted in \cite{need} and \cite{selection}, when used for binaural rendering, HRTFs produce the most accurate localization results when customized to an individual. The significance of personalized HRTFs for XR (Extended Reality including virtual reality (VR), augmented reality (AR) and mixed reality (MR)) and spatial audio applications is further emphasized in \cite{hybrid}, where it is stated that providing users with personalized HRTF sets is crucial for a truly immersive XR experience, free of localization errors and inside-the-head sound perception.
\medskip

\nomenclature{\(XR\)}{Extended reality}

Generalized HRTFs, while more accessible, lack the precision and realism offered by individualized HRTFs. The uniqueness of each individual's head morphology can lead to compromised results when using a generic HRTF for virtual acoustic display, resulting in diffuse or displaced auditory cues such as front-back confusion and elevation errors \cite{Global, Localisation}. To overcome these limitations, researchers have been exploring the benefits of obtaining individualized HRTFs from real human subjects, considering variations in head and ear anatomy which affect sound perception \cite{survey}. To address these challenges and unlock the full potential of spatial audio experiences, further research and implementation of individualized HRTFs is required.
\medskip

Indeed, recognizing the variations in individual head and ear anatomy and developing techniques to capture and utilize individualized HRTFs are decisive steps towards advancing spatial audio technologies and enhancing the realism of sound localization in virtual environments and beyond. For this reason, is is necessary to study the head anthropometry to fully understand the anatomical variation within the human diversity.

\section{Head anthropometry}

Ghorbal et al. \cite{pinnainfluence} stated that the HRTFs of a listener are intimately related to his morphology. Thus, a good knowledge of the listener's shape and anthropometry should be a relevant condition for inferring his HRTFs. The morphology of the human head and ears can be represented by measurements. These specific measurements allow one to have individualized anthropometric parameters \cite{hrtfmagn}.
\medskip

The study of the physical dimensions and characteristics of the human head, including its size, shape, and individual anatomical features allow for a repetitive pattern to appear and key points in terms of anthropometry of the head. Head anthropometry is a crucial aspect of HRTF, as it explains how sound interacts with an individual's head, ears, and torso, influencing the personalized acoustic filtering effects which determine sound localization. To create accurate and effective personalized HRTFs, it is essential to obtain precise anthropometric measurements of each individual. 
\medskip

Specifying a general set of well-defined and relevant measurements is problematic.The selection of head anthropometric measurements for creating standardized head and ear models is a result of research and collaboration among experts in the fields of audiology, acoustics, and spatial audio. A mutual agreement on the most relevant and significant dimensions which influence HRTFs and sound localization must be determined.
\medskip

The problem is particularly difficult for the pinna, where small variations can produce large changes in the HRTF. Anthropometric measurements, even if imperfect, enable the investigation of possible correspondences or correlations between physical dimensions and HRTF \cite{CIPIC}. Their study has led to the understanding of their physiognomy variations and their influences on Head-Related Transfer Functions. 
\medskip

Two standardized templates of head and ear measurements for average anthropometric models stand out in several research papers (referenced below). These models described in the next two sections, represent the average/typical head and ear shapes and serve as generalized templates when collecting anthropometric measurements. 
\medskip

\begin{figure}[!htbp]
     \centering
     \includegraphics[width=9cm]{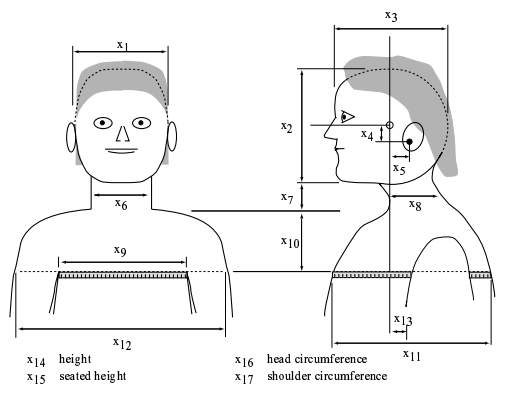}
     \caption{Head and torso measurements from CIPIC \cite{CIPIC}}
    \label{CIPIC head and torso}
    
\end{figure}

\subsection{CIPIC anthropometric measurements}
\nomenclature{\(CIPIC\)}{Centre for Image Processing and Integrated Computing}

In 2001, the Centre for Image Processing and Integrated Computing (CIPIC)\cite{CIPIC} of the University of California, Davis, publicly released a database providing anthropometric parameters that included 17 head and torso parameters and 10 pinna parameters. Statistics of anthropometric parameters and correlations between anthropometry and some temporal and spectral characteristics of HRTFs are specified in the CIPIC publication \cite{CIPIC}. The researchers sought a general behaviour which can be estimated from using simple geometric models of the torso, head and pinna. They wanted their models to be individualized to individual listeners when appropriate anthropometric measurements were available. 
\medskip

\begin{figure}[!htbp]
     \centering
     \includegraphics[width=8cm]{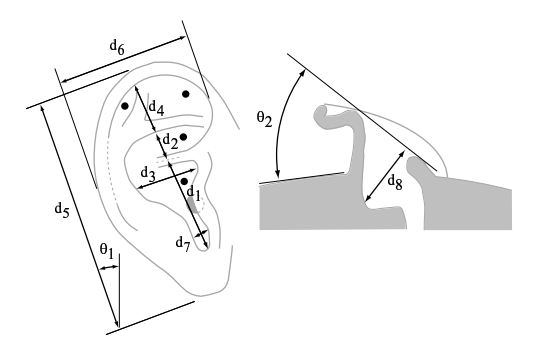}
     \caption{Pinna measurements from CIPIC \cite{CIPIC}}
    \label{CIPIC pinna}
    
\end{figure}

The choice of anthropometry relevant to understanding or estimating HRTFs led them to define a set of 27 anthropometric measurements. 17 for the head and torso and 10 for the pinna. These specific parameters are shown in figure \ref{CIPIC head and torso}, figure \ref{CIPIC pinna} and are listed in Table \ref{CPIC parameters}.
\medskip

\begin{table}[!htbp]
\centering
\begin{tabular}{ |c|c|c|c|}
\hline
         Variable  & Measurements & Variable  & Measurements \\
    \hline
     $x_1$ & head width  & $x_{15}$ & seated height\\
    \hline
    
     $x_2$ & head height & $x_{16}$ & head circumference\\
    \hline
     
     $x_3$ & head depth & $x_{17}$ & shoulder circumference\\
    \hline
    
     $x_4$ & pinna offset down & $d_1$ & cavum concha height\\
    \hline
     
     $x_5$ & pinna offset back & $d_2$ & cymba concha height\\
    \hline
    
     $x_6$ & neck width & $d_3$ & cavum concha width\\
    \hline
   
     $x_7$ & neck height & $d_4$ & fossa height\\
    \hline
  
     $x_8$ & neck depth & $d_5$ & pinna height\\
    \hline
     
     $x_9$ & torso top width & $d_6$ & pinna width\\
    \hline
 
     $x_{10}$ & torso top height & $d_7$ & intertragal incisure width\\
    \hline
    
     $x_{11}$ & torso top depth & $d_8$ & cavum concha depth\\
    \hline
  
     $x_{12}$ & shoulder width & $\theta_1$ & pinna rotation angle \\
    \hline

     $x_{13}$ & head offset forward & $\theta_1$ & pinna flare angle\\
    \hline
   
     $x_{14}$ & height && \\
    \hline
\end{tabular}
    \caption{Anthropometric variables from CIPIC \cite{CIPIC}}
    \label{CPIC parameters}
\end{table}

CIPIC was an early institution to conduct significant research in spatial audio and Head-Related Transfer Functions. Their work laid the foundation for understanding the importance of HRTFs in creating immersive sound localization experiences. The CIPIC anthropometric measurements have become a benchmark and have been used in numerous research relating to individualized HRTFs and spatial rendering techniques. CIPIC allowed the HRTF database to be accessible to the research community and developers. This open access approach encouraged widespread use and collaboration, contributing to its popularity and reputation. The following works have been using and mentioning the CIPIC database and based their work on their anthropometric models \cite{ITD3, Ziegelwanger, extracted, extraction cipic, Magnitude, sameproject, inoue, xu1, xu2, hugeng, Grijalva}. They will be presented in Section \ref{related}. 
\medskip

The database also contains the Head-Related impulse response (HRIR) of 45 subjects at 25 azimuths and 50 elevations. The spatial sampling is roughly evenly distributed on a sphere with a radius of 1m. The CIPIC database thus provide impulse responses of 45 subjects with their corresponding anthropometric measurements. This is an essential point for linking anthropometric features and their influence on HRTFs.

\subsection{HUTUBS anthropometric measurements} \label{hutubsection}

The HUTUBS HRTF database was created in 2019 at the Technical University of Berlin (TU Berlin) in the Audio Communication Group led by Fabian Brinkmann \cite{hutub}. A fusion between TU Berlin, Huawei and Sennheiser enabled the creation of a dataset which contains Head-Related Transfer Functions, 3D head meshes and anthropometric features on a range of 96 different subjects. The full overview of available data is shown in Figure \ref{hu_over}.
\medskip

\begin{figure}[!htbp]
     \centering
     \includegraphics[width=13cm]{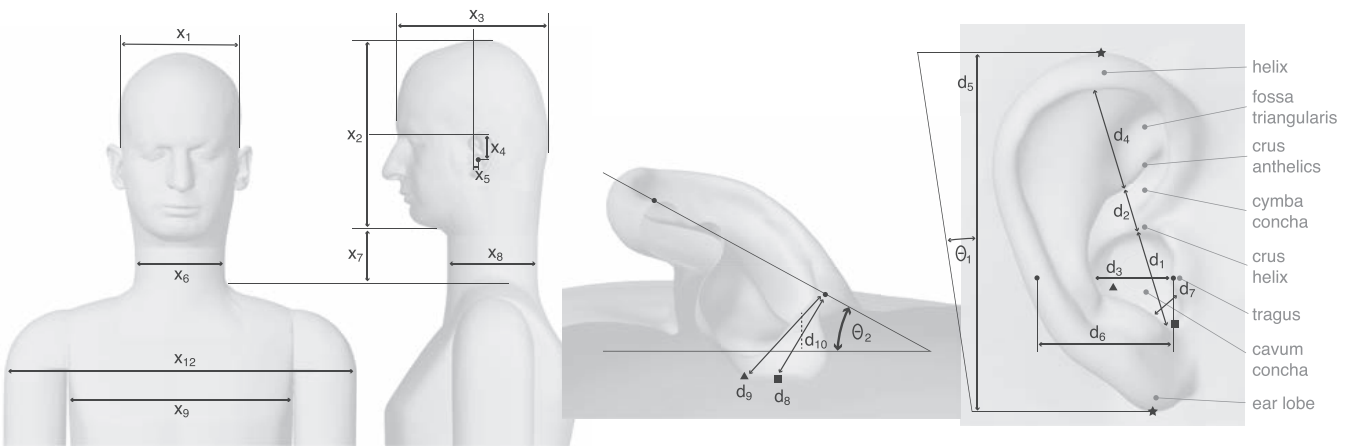}
     \caption{Definition of anthropometric measures from HUTUBS \cite{cross-eva}}
    \label{hutubs measures}
\end{figure}

HUTUBS provides a set of anthropometric measurements concerning body, head and pinna for only 93 subjects out of the 96. The set of measurements is similar to the one reported in CIPIC dataset \cite{CIPIC}, but it has some alterations and additions. The HUTUBS dataset has led to several improvements with respect to older datasets, such as CIPIC.
\medskip

HUTUBS has gathered 25 anthropometric features extracted from the 3D mesh. To eliminate the bias of manual measurements, the torso, head and pinna features were extracted fully automatically when possible by finding characteristic points on the mesh outline following the definition from Algazi et al. \cite{CIPIC}. In some cases, to ease the automatic extraction of the measurements or because some area were blurry (i.e. unclear), the original definitions were modified. This results in the accuracy of the measurements being inconsistent and the exact positions of the measuring points are not provided. The 3D meshes acquisition will be explained in Section \ref{mesh}.
\medskip

Figure \ref{hutubs measures} show the HUTUBS anthropometric measures definition and Table \ref{Hutubs parameters} list the different variables with their corresponding features.

\begin{table}[!htbp]
\centering
\begin{tabular}{ |c|c|c|c|}
\hline
         Variable  & Measurements & Variable  & Measurements \\
    \hline
     $x_1$ & head width  &  $d_1$ & cavum concha height\\
    \hline
    
     $x_2$ & head height & $d_2$ & cymba concha height\\
    \hline
     
     $x_3$ & head depth & $d_3$ & cavum concha width\\
    \hline
    
     $x_4$ & pinna offset down & $d_4$ & fossa height\\
    \hline
     
     $x_5$ & pinna offset back & $d_5$ & pinna height\\
    \hline
    
     $x_6$ & neck width & $d_6$ & pinna width\\
    \hline
   
     $x_7$ & neck height & $d_7$ & intertragal incisure width \\
    \hline
  
     $x_8$ & neck depth & $d_8$ & cavum concha depth (down)\\
    \hline
     
     $x_9$ & torso top width & $d_9$ & cavum concha depth (back) \\
    \hline
 
     $x_{12}$ & shoulder width & $d_{10}$ & crus helix slant height\\
    \hline
    
     $x_{14}$ & height & $\theta_1$ & pinna rotation angle\\
    \hline
  
     $x_{16}$ & head circumference & $\theta_2$ & pinna flare angle \\
    \hline

     $x_{17}$ & shoulder circumference & &\\
    \hline
   
\end{tabular}
    \caption{Anthropometric variables from HUTUBS \cite{cross-eva}}
    \label{Hutubs parameters}
\end{table}

As mentioned above, the HUTUBs database provides a total of 96 sets of Head-Related Impulse Responses (HRIR), 96 Headphone Impulse Responses, 93 sets of anthropometric measures and 58 head 3D meshes. Headphone impulse responses (HpIRs) refer to the acoustic responses recorded at the eardrums when headphones are worn. It captures the unique sound characteristics and spatial cues delivered by the headphones to the listener's ears. This wealth of data allows researchers to dive into the intricate relationship between individual ear anatomy and its acoustic impact on sound perception. The extensive number of subjects in the dataset ensures a diverse representation of human ear shapes, enabling thorough investigations into the variability of HRTFs across individuals.
\medskip

With the availability of anthropometric measurements, statistical analyses and correlations between ear anatomy and specific acoustic properties can be explored. These analyses offer valuable insights into the individualized nature of HRTFs and their impact on sound localization and spatial audio perception. The compatibility of the HUTUBS dataset with the SOFA format support widespread adoption and comparability across various research studies, facilitating collaboration and standardization within the research community.
\medskip

For all these attributes, the HUTUBS dataset have been widely used by researchers and published in scientific papers such as the following \cite{predict, extracted, selection, hybrid, Global, Pelzer, Deep learning}. The different approaches presented in those papers will be detailed in the next section.

\nomenclature{\(HpIRs\)}{Headphone Impulse Responses}

\section{HRTF individualization related works} \label{related}

Numerous research has been devoted to the domain of personalized Head-Related Transfer Functions, exploring diverse techniques such as individual measurements, numerical simulation, anthropometric extraction or HRTF selection from databases. Each approach offers distinct advantages and faces unique challenges. This chapter provides a comprehensive review of the published literature, their concept, approach as well as their methods.
\medskip

Individualized acoustic measurements for personalized HRTFs is the main way of offering tailored spatial audio experiences to the public, however it comes with significant complexities. The process of acquiring individual HRTFs measurements in an anechoic room with the use of calibrated speakers and microphones is time-consuming, labour-intensive and requires specialized equipment and expertise. In order to ease the acquisition of personalized HRTFs, a need for more modern techniques is required. 
\medskip

Although Ziegelwanger et al. and Katz et al.  \cite{Ziegelwanger, Katz} demonstrate that it is possible to numerically simulate individual HRTFs from a head mesh, simulation techniques require a precise 3D mesh of the subject and thus a large quantity of anthropometric measurements to reproduce the subject head and ears accurately. The simulation method takes a lot of computing time and is yet not automated. It is a laborious process which can take several hours per subject and is not easily accessible to the public.
\medskip

Simone Spagnol \cite{selection} proposes a method of using just a few anthropometric parameters of the head and torso to select with linear regression the most accurately fitting HRTF for a given subject from a set of non-individual HRTFs. His approach is especially attractive because it requires minimal effort on the user's behalf. The HUTUBS dataset was used to evaluate the proposed model for selecting the best fitting non-individual HRTF for a given subject. The results indicate that it can identify non-individual HRTFs with low localization error which was evaluated through subjective perceptual tests. However, anthropometric parameters need to be given from the subject and specific measurement points must be repected.
\medskip

\nomenclature{\(CNN\)}{Convolutional Neural Network}
\nomenclature{\(SH\)}{Spherical Harmonic}
\nomenclature{\(SHT\)}{Spherical Harmonic Transform}

Zhao et al. \cite{Magnitude}  propose a  personalized HRTF method based on anthropometric measurements and ear images of a subject. The model consists of two sub-networks. The first is a Convolutional Neural Network (CNN) (see Section \ref{CNN} model which extracts features from ear images. The second is a sub-network which uses anthropometric measurements given by the subject, ear features and frequency information to predict spherical harmonic (SH) \footnote{Spherical Harmonics (SH) are mathematical functions defined on the surface of a sphere used to represent spatial information and directional properties in 3D environments and audio simulations \cite{Magnitude}} coefficients. The personalized HRTF is obtained through inverse spherical harmonic transform (SHT) reconstruction. This method still requires anthropometric data that the subject has to provide. The document uses the CIPIC database and the AWE ear images database. The AWE database, published by the University of Ljubljana Slovenia, provides 1000 ear images of different subjects but does not include landmarks or key points. 
\medskip

Wang et al. \cite{Global} also propose an approach with spherical harmonics for global HRTF personalization employing subject anthropometric features using spherical harmonics transform and convolutional neural network. Using the HUTUBS HRTF database as a training set for a learning algorithm, a SHT was used to produce personalized HRTF’s for several spatial directions. This method is more complex in terms of mathematics and requires the computation of spherical harmonic equations in which details provided are insufficient and lack clarity to implement.
\medskip

\nomenclature{\(DNN\)}{Deep Neural Network}

Lee et al. \cite{DNN modeling} propose a personalized Head-Related Transfer Function estimation method based on deep neural networks by using anthropometric measurements and ear images. The proposed method consists of three sub-networks for representing personalized features and estimating the HRTF. As input features for neural networks, the anthropometric measurements regarding the head and torso are used for a Deep Neural Network (DNN)(see Section \ref{deeplearning}) and the ear images are used as inputs for a convolutional neural network (CNN). The outputs of these two sub-networks are merged into another DNN in order to estimate the personalized HRTF. 
\medskip

A further alternative consists of synthesizing HRTFs from a small number of anthropometric measurements of the listener’s head, ears, and/or torso. Bilinski et al. \cite{Bilinski} propose a method for HRTF synthesis using sparse representation which involves finding a sparse linear combination of basis vectors that best approximates the HRTF measurements for a given individual. Grijalva et al. \cite{Grijalva} present an anthropometry-based method to personnalize HRTF using multiple learning in both azimuth and elevation angles with a single nonlinear regression model (statistical approach to describe nonlinear relationship in the data).
\medskip

These methods appear to be attractive especially for their use of ear images which can be easily provided by the subject interested in personalized HRTFs, but they still require anthropometric measurements respecting specific points. 
\medskip

Riccardo Miccini and Simone Spagnol's \cite{Deep learning} paper focuses on HRTFs individualization using deep learning techniques and explore three different directions : supervised learning, unsupervised learning, and anthropometric data-based synthesis. The use of supervised learning techniques aims to extract predictors from user data, such as anthropometric measurements to estimate a subject's HRTF set. Unsupervised learning techniques such as autoencoder networks are studied to learn a compact representation of HRTFs which can be used to synthesize them. The authors approach an autoencoder as a feed-forward neural network composed of two subnets: an encoder network and a decoder network. The encoder network maps the input HRTF to a compact representation, and the decoder network maps the compact representation back to the original HRTF. Finally, synthesis of HRTFs from anthropometric data using deep multi-layer perceptrons (i.e. building blocks of neural networks, see Section \ref{deeplearning})  and principal component analysis\footnote{Principal Component Analysis (PCA) is a statistical technique used for reducing the dimensionality of data while retaining as much of its variability as possible \cite{Deep learning}} is explored. The use of publicly available HRTFs database like HUTUBS and machine learning techniques to choose, adapt, or estimate a subject's HRTF set based on anthropometric measurements is studied. This research paper lacks to provide enough information on the implementation of the different method and the source code to implement the different techniques is not available. Anthropometric measurements are required and no easily accessible way for individualized HRTFs to the public is offered.  
\medskip

In another paper from Riccardo Miccini and Simone Spagnol \cite{hybrid}, the authors present a hybrid approach to individualized Head-Related Transfer Function (HRTF) modeling which requires only 3 anthropometric measurements and an image of the pinna. A prediction algorithm based on variational autoencoders synthesizes a pinna-related response from the image, which is used as a filter to a measured response from the person's head and torso. This filter is then adjusted to match the interaural time difference of the HUTUBS dataset, which helps to minimize errors in sound localization. This method still requires 3 measurements of the subjects anthropometry but is getting closer to an easy and accessible way for the public to obtain personnalized HRTFs.
\medskip

Fantini et al. \cite{extracted} paper offers an HRTF individualization method based on anthropometric features which are automatically extracted from 3D head meshes, their acquisition is explained in Section \ref{mesh}. The process is articulated in two different phases: a pinna anthropometry extraction and an HRTF individualization. In the pinna anthropometry extraction phase, a set of anthropometric parameters related to pinna is measured on the 3D mesh. In the second phase, the relationship between pinna anthropometry and HRTFs based on the HUTUBS dataset is modeled. For each elevation angle considered, a Regression Neural Network is trained to predict the corresponding HRTF, given the anthropometry. The method aims at a fully automated process which is able to estimate individualized HRTF starting from a 3D mesh of the subject’s pinna.
\medskip

Dinakaran et al. \cite{extraction cipic} evaluate the precision of anthropometric features which are automatically extracted from individual 3D head-and-shoulder meshes. These meshes are generated with a low-cost 3D scanning device (the Kinect sensor), by identifying and measuring distances between characteristic points on the outline of each mesh. A comprehensive set of anthropometric features was automatically extracted for 61 subjects. By cross validation with the manually extracted values, the method was found to yield accurate and reliable estimations of corresponding features. It is stated in the paper that manual work was needed to process the 3D meshes in order to be ready for the extraction step. 3D meshes require more resources to obtain anthropometric features. Wang et al. \cite{predict} also used 3D meshes resulting in scanned head geometry for HRTF personalization employing CNN. Their neural network inputs are head and torso measurements, spherical harmonic coefficients, a frequency target and 3D ear meshes. All these inputs make the end-user experience challenging.
\medskip

These research do not provide any source code for implementing or further information on the algorithms and equations used. It is also stated that manual intervention is needed to correctly process the 3D mesh. The acquisition of the 3D mesh requires specific devices and methods such as laser scanning or image reconstruction from pictures of different angles. 
\medskip

The extraction of anthropometric measurements from ear images seems achievable and more accessible to implement. The concern is to be certain that ear anthropometric features are relevant enough to offer personalized HRTFs. This concern will be addressed by reviewed several published papers that only retained ear anthropometric features. 
\medskip

The following documents propose methods for individualized HRTFs using anthropometric measurements. They performed correlation analysis between anthropometric parameters within the CIPIC database. They researched which parameters were the most relevant for suitable HRTFs selected from anthropometric parameters. Inoue et al. \cite{inoue}, Xu et al. \cite{xu1, xu2} and Hugeng et al. \cite{hugeng} all retained measurements that were related to the ear. 
\medskip

The goal of this Master's Thesis is not specifically to achieve a high quality tailored HRTF for everyone for a but to make the individualization of HRTFs accessible to a large range of people. Focusing on ear anthropometric to be able to use ear images and extract chosen anthropometric features with the use of deep learning is a reasonable choice for a project which is exploring the individualizaion of HRTFs using artificial intelligence technology. 
\medskip

Zotkin et al. \cite{sameproject} proposed an HRTF personalization algorithm based on digital images of the ear taken by a video camera. They performed 7 measurements on it with manually annotated landmarks and computed out of them a distance between subjects of a given database. The closest match was selected and the HRTFs are used as raw material for the individualisation experiment. Zhang et al. \cite{ITD3} also explore the option of finding an HRTF which is the most compatible in the CIPIC ear picture database based on pinna features extracted from profile pictures using specific filters on the image.
\smallskip

Geronazzo et al. \cite{pinna extraction} present a system for customized binaural audio based on the extraction of the relevant features from a 2-D representation of the listener’s pinna. A procedure based on multi-flash imaging for recognizing the main contours of the pinna and their position with respect to the ear canal entrance is detailed. Their process is only presented and no source code or algorithms have been made available to implement their work. 
\smallskip

With the lack of available source code or repositories offering the implementation of personalized HRTFs, the following work explores the option to extract anthropometric measurements from ear pictures using a trained Convolutions Neural Network. These measurements will be used to form a 7 dimensional vector of distances in order to perform a best match identification within an existing HRTFs database.

\newpage
\chapter{HRTF individualization algorithm} \label{method}

While every anatomical head component shapes the incoming sound waves, this work is focused on the most individual one, the pinna.  However, instead of directly measuring anthropometric data from the pinna, the approach presented in this work aims to automatically extract this data from an image of the subjects ear.
\medskip

Simone Spagnol stated in his work on HRTF Selection by Anthropometric measurements \cite{selection}, "The use of a small number of anthropometric parameters for estimating the relative fitness of a non-individual HRTF set compared to another is especially attractive because of the little effort it takes on the user’s behalf". Making an easy way for a user to obtain his personnalized set of HRTFs by only using an image of his pinna is the objective in order to reduce the effort on the user’s behalf.
\medskip

This approach was first pursued by Zotkin et al. \cite{rendering} who proposed to select the HRTF set which best matches an anthropometric data vector of the pinna. Manual positioning of the measurement points had to be selected for each subject, and no automation was implemented. 

\bigskip

\begin{figure}[!htbp]
     \centering
     \includegraphics[width=15cm]{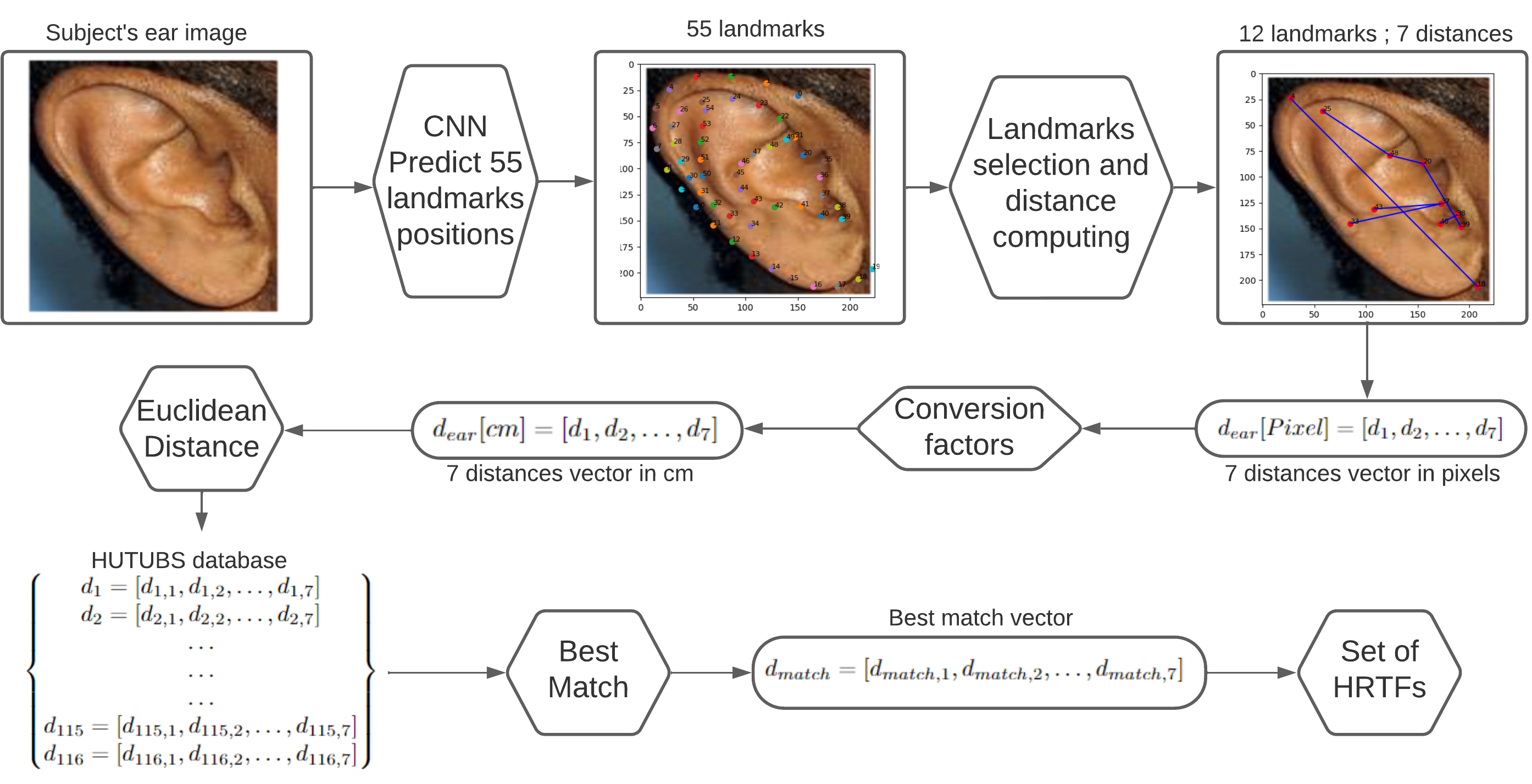}
     \caption{Scheme of the HRTF individualization process}
    \label{scheme}
\end{figure}
\bigskip

The use of a CNN will automate the process of extracting measurement points from images of subjects ear. The CNN will be trained on a large quantity of ear images with corresponding 55 specific landmarks (i.e. reference points on an object). These landmarks define the biometric characteristics of the pinna. Following specialized training, the CNN will be able to place 55 landmarks in an image of the ear. 12 relevant landmarks, referring to 7 anthropometric measurements established by the HUTUBS database, will be selected. An algorithm will compute the 7 distances in pixels from the 12 landmarks positions. 
\medskip

These computed distances will form a 7 distances vector in pixels and will be converted into centimetres using conversion factors. A best matching method will then be applied to find the closest vector in the HUTUBS database (116 subjects) of anthropometric measurements using the Euclidean distance equation. The database also includes the corresponding set of HRTFs for each set of 7 anthropometric measurements. The best match of HRTFs will be found for the distances extracted from the subject's ear image within the available set of HRTFs provided by the database. The various steps of the HRTF individualization process are represented in a scheme in Figure \ref{scheme}. 
\medskip

The methodology will be executed in multiple steps, starting with the training of the CNN on a dataset of ear images and landmarks. A test phase will then follow to evaluate the ability of the CNN to position landmarks on an unseen ear image. Distances will then be computed based on the appropriate landmarks positions, and finally, the best matching distance vector will be selected from the available anthropometric measurements database, thereby identifying the optimal set of HRTFs for the subject's ear characteristics.

\section{Ear landmarks extraction}

In order to place specific landmarks on the ear, a neural network model must be designed, trained and tested. A learning model which can predict landmarks correctly on the shape of the ear following its characteristics which is a useful tool for automation. When trained, the model would be able to place landmarks on unseen images provided by the subjects and thus be effective on a large number of people. 
\medskip

A convolutional neural network has been chosen for its ability to automatically learn relevant hierarchical features from data and capturing local patterns. CNNs have been widely used in image-related tasks and have remarkable success in various computer vision tasks including object recognition, image segmentation, and landmark detection. The principle of deep learning and neural networks is explained in Section \ref{deeplearning}. The nature and mechanisims of CNNs are detailed in Section \ref{CNN} as well as details on the training and test phases.
\medskip

The inherent convolutional layers help the network focus on local image regions, and through a training process, the model learns to identify distinctive patterns and spatial arrangements in the input images, enabling landmark positioning. CNNs can handle variations in scale, rotation, and perspective within the images, making them robust to different viewing conditions. As the ear picture of a subject could vary in color, brightness, framing and orientations, having a robust trained model is crucial for the diversity of data input. 
\medskip

This procedure requires the acquisition of a database which is split in two different sections. The training section contains the images and the corresponding annotated landmarks on which the model will be trained on. The test section will contain images unknown to the model on which the model can be evaluated for its performance. The dataset must contain a large amount of labelled samples to avoid over-fitting problems\footnote{\label{overfitting}Over-fitting occurs when a machine learning model becomes overly specialized to the training data. The model then captures noise and irrelevant patterns leading to reduced generalization performance on new unseen data \cite{Magnitude}}. Data augmentation techniques will therefore be implemented to increase the number of training samples. 
\medskip

An available repository on GitHub \cite{ear-landmark} provides a CNN architecture as well as a dataset which includes ear images and manually annotated landmarks. The repository also includes a trained model on the data provided which avoids the time-consuming and resource-intensive nature of training. The CNN and its trained model will be implemented and tested with the provided dataset. The result will be analyzed and discussed for potential use in HRTF individualization.

\subsection{Landmarks}
\nomenclature{\(I-BUG\)}{Intelligent Behaviour Understanding Group} 

The GitHub repository \cite{ear-landmark} refers the original data to be provided by I-BUG ( Intelligent Behaviour Understanding Group) from the Department of Computing from the Imperial College in London who offer Deformable Models of Ears In-The-Wild for Alignment and Recognition \cite{ibug}. I-BUG refer to the importance of the ear within its biometric which has gained significant attention in the field of automatic identity verification. It is in this area of interest of being able to find the identity of someone with biometric characteristics of its ear that the database has been crafted and assembled. The distinctive features of the outer ear, including the outer helix, anti-helix, lobe, tragus, and more, make it a valuable biometric characteristic \cite{Zhou}.
\medskip

\begin{figure}[!htbp]
     \centering
     \includegraphics[width=7cm]{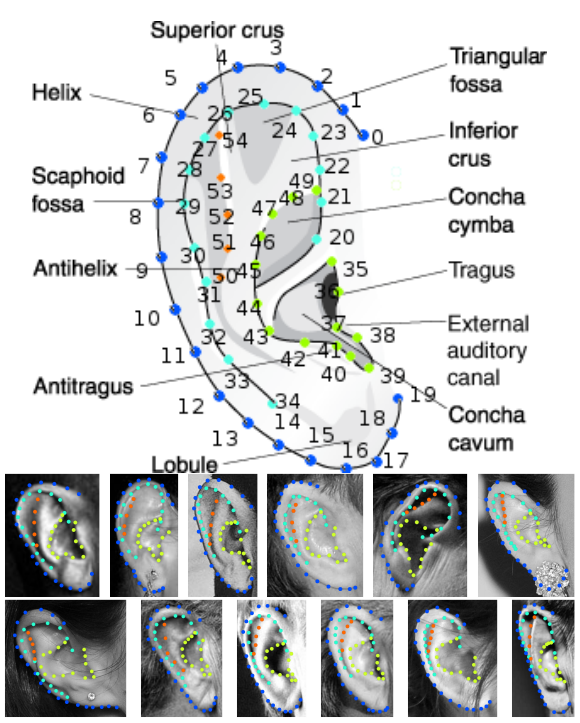}
     \caption{I-BUG : Example of the annotated 55 landmarks on ears \cite{ibug}}
    \label{ibug}
\end{figure}

Alignment plays a crucial role in ear recognition, as the ear is a deformable object. I-BUG address this alignment by providing a landmark-annotated "in-the-wild" ear database. "In-the-wild" images refers to pictures which have been collected on Google Image with no specific identity, by searching using the ear related tags. Each image is manually annotated with 55 landmarks points. The framework of the pinna anatomy on which the landmarks positioning is based as well as examples of annotated images are shown on Figure \ref{ibug}.
\medskip

The semantic meaning of the 55 landmarks are: 
\begin{itemize}
    \item Ascending helix [0-3]
    \item Descending helix [4-7]
    \item Helix [8-13]
    \item Ear lobe [14-19]
    \item Ascending inner helix [20-24]
    \item Descending inner helix [25- 28]
    \item Inner helix [29-34]
    \item Tragus [35-38]
    \item Canal [39]
    \item Antitragus [40-42]
    \item Concha [43-46]
    \item Inferio crus [47-49]
    \item Supperio crus [50-54]
\end{itemize}

\subsection{I-BUG dataset}
The dataset consists of 500 training images, 105 test images and corresponding landmarks for each image. The picture contains the full human body of the person so a re-framing of the images with a focus on the ear has to be applied using a python script provided by the repository. The dataset is thus assembled with the re-framed images and the corresponding landmarks. The size of the images is 224 pixels by 224 pixels which form a square image of 224x224 pixels. Example images are represented in Figure \ref{8}. 

\medskip

In order to have a large amount of training data for the neural network training, the data must be augmented \cite{dataaugmented}. To achieve that, modifications to the data can be applied to enlarge the available dataset. The data augmentation is performed by several modifications made on the image and are the following :
\medskip

\begin{enumerate}
    \item Vertical flip
    \item Left rotation
    \item Right rotation
    \item Vertical flip and Left rotation
    \item Vertical flip and Right rotation
\end{enumerate}

These modifications enlarge the dataset of ear images to 3000 training images and 630 test images. The same procedure is applied on the landmarks to maintain the duality between the images and the landmarks.

\subsection{Neural network architecture} \label{neural network}

The neural network proposed on the GitHub \cite{ear-landmark} is a CNN (Convolution Neural Network) which is a deep learning algorithm designed for processing and analyzing visual data such as images or videos. It is particularly effective in capturing complex patterns and features from the input data. The key components of a CNN are convolutions layers, pooling layers, and fully connected layers. CNN usual vocabulary will be used in the following sections but a complete explanation of machine learning and CNNs, the operating mechanisms of neural networks is addressed in Section \ref{deeplearning} as well as the CNN concept and usual layers in Section \ref{CNN}. 
\medskip

The model architecture can be described in 6 main components : 
\medskip
\begin{itemize}
    \item The 'Conv2D' layers perform convolutional operations on the input images. Convolutional layers apply filters (small matrices) to the input data, enabling the network to extract local patterns and features. These filters detect edges, shapes, textures, and other visual elements. Each 'Conv2D' layer applies a set of filters to extract specific features from the input. The numbers 16, 32, 64, 128, 256, and 512 represent the number of filters in each layer. The (3, 3) and (5, 5) indicate the size of the filters.

    \item The 'MaxPooling2D' layers perform down-sampling which reduces the size of the images while retaining the most essential information by selecting the maximum value with a specific pool size. A pool size determines the size of the local regions over which the downsampling is performed. It reduces the spatial dimensions of the feature maps, helping to capture important features while reducing computational complexity because of the inferior sizes.

    \item The 'BatchNormalization' layers normalize the outputs of the previous layers to have zero mean. This signify that the average value of the data is subtracted from each data point, effectively shifting the data distribution closer to the origin which is zero. It ensures that the data flowing through the network is centered around zero which allow more stable and efficient training. It normalizes the output also to have a unit variance, ensuring that the data has a consistent spread or dispersion which helps in maintaining stable gradients during training. By normalizing to a unit variance, the network becomes less sensitive to the scale of the data and can adapt more efficiently to different types of inputs. These two actions ensure that the network trains faster and is more stable. This normalization ensures that the data is centered around zero with a standard deviation of one, creating a more standardized and stable input distribution for each layer of the neural network. It also contributes to improving the convergence and generalization which allows the model to increase accuracy on new data which it has never encountered before during training, indicating its ability to capture underlying patterns rather than memorizing specific examples.

    \item The 'Dropout' layers randomly deactivate a specified fraction of the neurons during training, which helps in preventing over-fitting by reducing the interdependence of neurons. By randomly dropping out neurons, the model is forced to rely on different sets of neurons for each input, which prevents the network from becoming overly dependent on specific neurons and features. Using 'Dropout' layers helps to reduce the risk of over-fitting and increases the robustness of the model, leading to greater generalization on unseen data.

    \item The 'Flatten' layer reshapes the multidimensional output of previous layers into a one-dimensional vector, preparing for the fully connected layers. The 'Dense' layers are fully connected layers which connects all the neurons from the previous layer to the next layer. Each Dense layer applies a specified number of neurons (example : 1024) and an activation function (e.g. Relu, see Section \ref{deeplearning}). The final Dense layer has 110 neurons, which indicates the output dimension for the classification task. Each landmark requires a set of two coordinates to be placed on an image. The 110 output neurons of the CNN are the 55 landmarks X and Y coordinates. X and Y represent respectively the horizontal and vertical pixels axis of the image.

\end{itemize}
To go further in detail, each layer of the neural networks is explained in Section \ref{layer} and shown in Figure \ref{model}. 
\medskip

In a neural network model, parameters are values which are learned during the training process and are used to make predictions. They represent the weights and biases of the model's layers as explained in Section \ref{deeplearning}. In the proposed model, there are a total of 9 033 230 parameters responsible for capturing the patterns and relationships in the input data of which 9 029 902 are trainable parameters. The remaining 3,328 parameters are fixed  and non-trainable, they don't get updated during the training process. They are associated with specific operations or layers in the model which have predefined functionality. These parameters are used for normalization and regularization\footnote{A regularization task in neural networks aims to prevent over-fitting and improve generalization by introducing constraints or penalties on the model's complexity during training like dropout and batch normalization layers \cite{Nuti}} tasks and are not adjusted based on the input data. 

\subsubsection{Training the CNN}

Once the dataset is ready to feed the neural network and the architecture is properly designed for the specific task, the training stage can then be implemented. The training process of this neural network involves iteratively updating its parameters using an optimizer to minimize the error between the predicted outputs (i.e. predicted coordinates of landmarks on the ear image) and the actual target values (i.e. true coordinates of the annotated landmarks) , thus enabling the network to learn from the data provided. The training process of the model is as follows : 

\nomenclature{\(MSE\)}{Mean Squared Error}
\nomenclature{\(2D\)}{2-Dimensional}
\nomenclature{\(3D\)}{3-Dimensional}

\begin{itemize}
    \item It begins by initializing the Adam optimizer (see Section \ref{deeplearning}) which is a learning rate optimization algorithm commonly used for training neural networks. The learning rate is chosen as 0.001, representing the step size at each iteration during the training process. $Beta_1 = 0.9$ and $Beta_2 = 0.999$ represent the exponential decay rates for the mean (first moment) and the uncentered variance (second moment). These two moments control the weighting of previous gradients and squared gradients when calculating the adaptive learning rate. The learning rate decay value is chosen as zero, so that the learning rate does not gradually reduce over each update.
    \item The model is then compiled, it configures the model for training by the optimizer, loss function, and metrics. In this case, the optimizer is set to Adam, the loss function is mean squared error (MSE) as explained in Section \ref{deeplearning}, and the metric used for evaluation is accuracy. The MSE loss is a common choice for regression problems where the goal is to minimize the difference between predicted and actual values.
    \item Finally, the model is trained by iteratively updating the model's parameters. It feeds the model with batches of input images and their corresponding target landmarks coordinates values, adjusting the weights and biases to minimize the loss function. The model is trained for 300 epochs (iterations) and the batch size of 64 determines the number of samples processed before updating the parameters. An epoch refers to one pass through the entire training dataset.
    
\end{itemize}

The training process returns a history object which contains information about the training process. It provides a record of the loss and accuracy values at each epoch, which is useful for evaluating the model's performance and visualizing the training progress. It is used to analyze the model's convergence, identify potential over-fitting or under-fitting, and make informed decisions about further training or fine-tuning. The model provided by the GitHub being already trained, this history object is not available. 
\medskip

The training process ends by saving the obtained model in a .h5 file also known as an HDF5 file standing for Hierarchical Data Format version 5. It provides a flexible and efficient way to store and organize large amounts of numerical data and associated metadata. The .h5 file store the model's architecture, weights, optimizer state, and other configuration details enabling easy reloading without retraining. The provided trained model is stored in a .h5 file and available for direct use.

\subsubsection{Testing the CNN}

The purpose of the test phase is to evaluate the performance of the model on the test data. The evaluation is performed and computes the loss and accuracy of the model. It takes the trained model, test samples, and their corresponding labels as inputs, then computes the loss and accuracy of the model on the test data. The loss value indicates the measure of error between the predicted and true landmarks, while the test accuracy represents the percentage of correctly classified samples in the test dataset. This test process assesses the effectiveness of the trained model in making predictions on unseen data and provides valuable performance metrics for evaluation.
\medskip

The obtained loss value is 0.0048, which indicates  the average discrepancy between the predicted landmarks of the model and the true landmarks of the test dataset. A lower loss value is desirable as it demonstrates that the model's predictions are closer to the true landmarks implying an improved fit and performance. By minimizing the loss during the training process, the model aims to improve its ability to accurately predict the landmarks on unseen data from the test dataset.
\medskip

The test accuracy obtained is 0.3365, indicating that the model correctly predicted the result for approximately 33. 65 \% of the test data. It suggests that the model's performance is relatively low as it is making accurate predictions for approximately one-third of the test data.

\subsubsection{Results and discussion} \label{55}

\begin{figure}[!htbp]
     \centering
     \includegraphics[width=8cm]{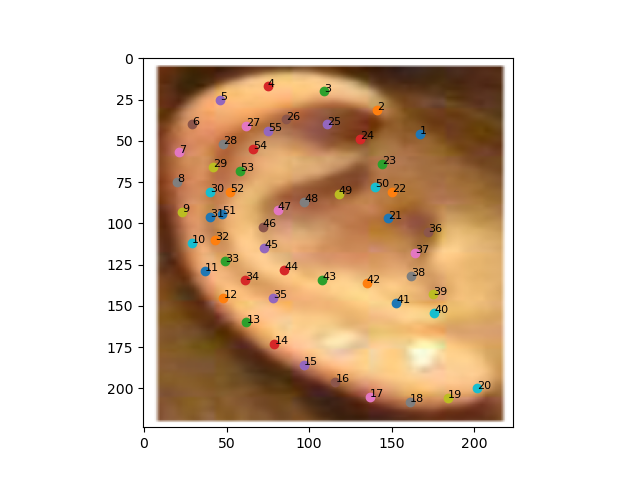}
     \caption{Image of the ear with the 55 predicted landmarks in color dots with their corresponding label}
    \label{exampleresult}
\end{figure}

In the result image presented in Figure \ref{exampleresult}, the labelled 55 landmarks on the ear are shown, demonstrating the model's capability to predict the locations of key points. However, the model achieved an accuracy of 33.65 \% on the test data, indicating room for improvement in the precise placement of landmarks. While the current performance may not be perfect, it showcases the fundamental functionality of the model, especially considering that it was only trained once with the provided dataset. Further enhancements can be achieved by training the model on a more extensive dataset comprising of a broader variety of images with corresponding landmarks, allowing the model to learn more robust and accurate representations of ear biometric characteristics. A specific metric would have to be introduced to compute the accuracy of the model to predict points and the error that occur for each result. This metric was not implemented in this work as the aim is to focus on feasibility rather than precision. Additional images of the CNN output are presented in Figure \ref{3_55}.
\medskip

The model has not been re-trained because of the validity of the database itself in terms of the end use for which this neural network is intended. These reasons will be explained in Section \ref{further}. The crucial point was to check the ability of a neural network to position 55 landmarks related to the ear shape to ensure that they were visually correct. Moreover, as explained in the following section, the 55 landmarks will be reduced in order to select relevant points which can be useful for the next steps of HRTF individualization.

\section{Relevant landmarks selection} \label{12}

By focusing on the biometric characteristic of the pinna, the HUTUBS database selected 12 ear anthropometric variables, see Table \ref{Hutubs parameters}. These variables represent the ear anthropometric measurements for each subject in order to individualize their anatomical differences, see Figure \ref{hutubs measures}. When observing a frontal image of the ear, only 7 of the anthropometric measurements are retrievable. These 7 variables will represent the biometric characteristics of an individual pinna in this work. 
\medskip

A similar choice was made by Zotkin et al. \cite{sameproject} in 2003, the authors selected 7 measurements following the CIPIC \cite{CIPIC} anatomic ear features. They took a digital image of the ear of an individual, manually annotated the landmarks, computed the distances, identified the anatomic features and found the best match in the CIPIC database. Their results indicate that this approach improves both the accuracy of the localization and the subjective perception of the virtual auditory scene. The generic trend which is observed is that incorporation of the KEMAR  model (see Section \ref{HAT}) almost always improves localization performance, whereas the ear
parameters based on the personalization method does not always perform well. The difference in this work is that the database used will contain more subjects than the CIPIC database and the anthropometric measurements will be automatically extracted and no manual intervention is required. 
\medskip

As detailed in Section \ref{55}, 55 landmarks are obtained on an ear image. Since not all the 55 points are of interest, 12 points are selected to represent the 7 anthropometric measurements corresponding to the HUTUBS definition represented in Figure \ref{hutubs measures}. From Figure \ref{ibug}, the following points have been chosen to represent the 7 distances. 

\begin{table}[!htbp]
\centering
\begin{tabular}{ |c|c|c|c|}
\hline
         Variable  & Measurements & Point A & Point B \\
    \hline

     $d_1$ & cavum concha height & 20 & 39\\
    \hline
     
     $d_2$ & cymba concha height & 20 & 48\\
    \hline
    
     $d_3$ & cavum concha width & 37 & 43\\
    \hline
   
      $d_4$ & fossa height & 25 & 48\\
    \hline
  
    $d_5$ & pinna height & 4 & 18\\
    \hline
     
     $d_6$ & pinna width & 33 & 37\\
    \hline
 
      $d_7$ & intertragal incisure width & 38 & 40\\
    \hline
    
\end{tabular}
    \caption{Chosen landmark points for the 7 distances}
    \label{landmarkchosen}
\end{table}

\begin{figure}[!htbp]
     \centering
     \includegraphics[width=7cm]{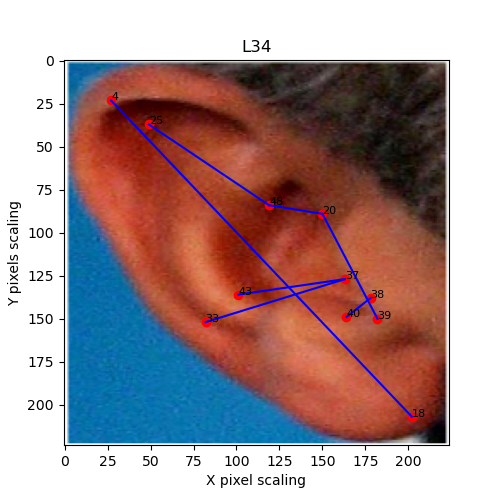}
     \caption{Image of the ear with the chosen predicted landmarks in red dots and connected with their corresponding distances in blue}
    \label{exampleland}
\end{figure}

The chosen landmarks can be selected from the 55, their coordinates retrieved and then displayed on the ear with red dots as well as the corresponding distances with blue lines. Figure \ref{exampleland} is an example of the chosen landmarks displayed on the input ear image of the model connected with the corresponding distances. Additional images are presented in Figure \ref{3_12}.
\medskip

\section{Distances computation}

The 12 landmarks positioned on the image, 24 coordinates ranging from 0 to 224 in pixels are used to compute the 7 anthropometric measurements. With $d_i$ having ($x_{iA}$, $y_{iA}$) and ($x_{iB}$, $y_{iB}$), the $d_i$ distance can be computed using the Euclidean distance with the following equation : 

\begin{equation} \label{equationdist}
    d_i = \sqrt{(x_{iB}-x_{iA})^2+(y_{iB}-y_{iA})^2}
\end{equation}
With $i = 1,2,...,7.$
\medskip

The 7 distances are thus computed in pixels scale from the output coordinates of the CNN. They are normalized (all divided by the same value) with the diagonal of the 224x224 square which is rounded to 316 pixels as it is the maximum distance that could be obtained with this image format providing distances varying between 0 and 1. To convert the distances into centimetres, a conversion factor for a distance computed in pixels in a 224x224 pixels image is required. The distances will only be relevant when they can be represented in centimetres, then it will be possible to find a best match of the 7 distances in an available database for an HRTF individualization.
\medskip

The HUTUBS database provides anthropometric measurements in pairs with HRTF sets of 93 subjects. The 7 distances selected in this work are stored in centimetres in an Excel file for each of the subjects right and left ear as well as the other 18 more anthropometric features. Once the distances in pixels are converted into centimetres, a best match method will be implemented to find, within the 93 subjects of the HUTUBS dataset, a set of HRTFs.
\medskip

The next section will explore a way of obtaining conversion factors using the HUTUBS 3D meshes. More specifically a zoomed image, resized in 224x224 pixels, of the ear from the 3D model will be obtained. Landmarks will be annotated to retrieve the 7 distances of the subjects in pixels and compare them with the given ones in centimetres with the aim of finding a conversion factor. 

\subsection{Conversion factor from HUTUBS 3D meshes and anthropometric data} \label{mesh}

\subsubsection{HUTUBS 3D head meshes}

\begin{figure}[!htbp]
     \centering
     \includegraphics[width=6cm]{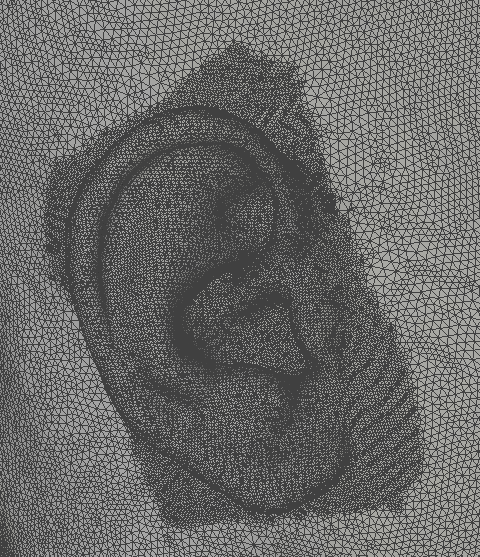}
     \caption{Zoom on HUTUBS subject 2 3D head mesh using Meshlab}
    \label{subject2zoom}
\end{figure}

The HUTUBS dataset contains 58 3D head meshes of different subjects. They were acquired using a Kinect 3D scanner for the head and a higher resolution Artec Space Spider scanner for the ear. The subject was seated on a swivel chair in a natural sitting position, the Xbox Kinect 3D scanner was positioned at a distance of 1 meter from the subject at eye level. A higher resolution surface scan with 0.05 mm point spacing resolution from a scanning distance of 0.2 m to 0.3 m was obtained for the left and right pinna using the Artec Space Spider \footnote{Artec3D.com : Artec Space Spider 2023. Consulted in May 2023. \url{https://www.artec3d.com/fr/portable-3d-scanners/artec-spider}}. The mesh density difference can be seen on figure \ref{subject2zoom} as the region around the ear has more points for a higher resolution. The subjects were wearing swimming caps to reveal the head shape and reduce the influence of hair on the scans. 
\medskip

\begin{figure}[!htbp]
     \centering
     \includegraphics[width=5cm]{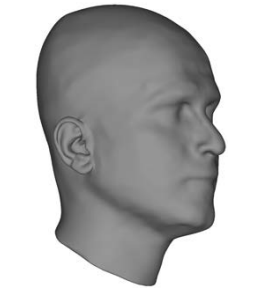}
     \caption{3D head mesh of subject number 2 from HUTUBS \cite{cross-eva}}
    \label{subject2}
\end{figure}

Once the scanning was completed, post-processing was applied to several head meshes to remove unwanted parts and any disturbance that could affect the head shape such as irregular parts close to holes in the mesh. The shoulders and torso were purposely removed by cutting the mesh at the bottom of the neck. The 3D holes in the scans were sealed using Artec Stutio \footnote{Artec3D.com : Artec Studio 17 2023. Consulted in May 2023. \url{https://www.artec3d.com/3d-software/artec-studio}} and the post-processing on the 3D Mesh were realized on Meshlab \footnote{Meshlab.com. Consulted in May 2023. \url{https://www.meshlab.net/}}. Geomagic point based glue tool was used to merge the two 3D scans and thus replace the pinna mesh of the Kinnect 3D by the Artec Space Spider. The final meshes were then aligned to the global coordinate system based on the interaural axis, defined as the axis connecting the centres of the entrances of the ear canals. An example of a subject 3D head mesh is shown on Figure \ref{subject2}. A closer look at the 3D head mesh is shown on Figure \ref{subject2zoom} highlighting the higher density of points for the ear mesh.

\subsubsection{From 3D head meshes to 2D ear images}

The 3D head meshes can be processed with python using the library Trimesh \footnote{Trimesh.org : Trimesh 3.22.1. Consulted in May 2023. \url{https://trimsh.org/index.html}} and Pyrender \footnote{Pyrender.com : Pyrender Documentation. Consulted in May 2023. \url{https://pyrender.readthedocs.io/en/latest/}}. Trimesh enables to read the 3D meshes provided by HUTUBS in ‘.stl’ format and load them as Trimesh objects. Pyrender allows to create a scene from the Trimesh object, add a perspective view and a directional light perpendicular to the ear for increased lighting. 
\medskip

\begin{figure}[!htbp]
     \centering
     \includegraphics[width=7cm]{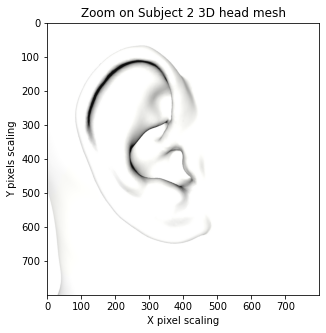}
     \caption{Zoom on HUTUBS subject 2 left ear in Python}
    \label{subject2pythzoom}
\end{figure}

Rotation is applied to obtain the correct side of the 3D head mesh. Then a zoom factor is used to obtain a closer look at the ear. A picture of the 3D ear model, as shown in Figure \ref{subject2pythzoom}, is taken for each side of the head. Additional ear images which were obtained with this technique are shown in Figure \ref{hu_cadre}. The left ear image is then mirrored in order to always have the same orientation of the ear. 

A loop is used to load every 3D meshes in consecutive order and the same procedure of rotation and image capturing is followed in order to obtain 116 images of the ear from the 58 3D head mesh as they contain two ears each. This procedure allows to obtain ear images of the 58 subjects which are not provided by HUTUBS. The 116 ear images and their corresponding anthropometric measurements constitute a unique dataset as no other dataset of this type is available. Even if the ears do not have their natural colours, the images can still be used to retrieve distances in pixels and be compared to distances in centimetres. In order to obtain the distances, the specific landmarks must be annotated. 

\subsubsection{Landmarks on HUTUBS ear images}

\nomenclature{\(HTML\)}{Hypertext Markup Language}
\nomenclature{\(CSS\)}{Cascading Style Sheets}

Once the 116 images are obtained, they can be resized to a 224x224 pixels size. As the HUTUBS dataset provides anthropometric measurements of the ears, the specific landmarks can be set on the ear images in order to retrieve the distances. An annotation tool available on GitHub \cite{gitlandmark} which is running on Python HTML (Hypertext Markup Language) and CSS (Cascading Style Sheets) enables to specify a name for a landmark and click on the location where the user wants to put the landmark. HTML is a standard language used for creating and structuring content on the web, defining the elements and layout of web pages. CSS is a stylesheet language used to define the visual presentation and layout of HTML documents. The combination of the two languages within Python libraries allow to use and interact with an annotation interface on Google Chrome. The code provided by the GitHub repository is modified and configured to annotate 12 landmarks with specific label on the 116 ear images retrieved from the HUTUBS head meshes. 

\medskip

The 12 landmarks are the locations where the distances in HUTUBS dataset would have been acquired. The locations are therefore assumed positions, as the actual positions are not supplied by the HUTUBS dataset. These position points are however estimated on the basis of a general ear diagram provided as shown on Figure \ref{hutubs measures} and explained in Section \ref{hutubsection}. The 12 landmarks positioned are used to represent the following distances listed in Table \ref{Hutubs features} and are the same as the ones selected in Section \ref{12}.

\begin{table}[!htbp]
\centering
\begin{tabular}{ |c|c|}
\hline
         Variable  & Measurements \\
    \hline

     $d_1$ & cavum concha height\\
    \hline
     
     $d_2$ & cymba concha height\\
    \hline
    
     $d_3$ & cavum concha width\\
    \hline
   
      $d_4$ & fossa height\\
    \hline
  
    $d_5$ & pinna height\\
    \hline
     
     $d_6$ & pinna width\\
    \hline
 
      $d_7$ & intertragal incisure width\\
    \hline
    
\end{tabular}
    \caption{Chosen anthropometric measurements}
    \label{Hutubs features}
\end{table}

The positions of these specific landmarks on the 224x224 pixels image are stored in individual '.json' files for each landmark before being converted in '.txt' files containing all the landmarks for each image. Therefore, the X and Y coordinates for the 12 landmarks of each ear within the 116 images are obtained and stored correctly for later use in the next step. 

\subsubsection{HUTUBS conversion factor}

A Python script is developed to read all the .txt files and ear images, retrieve the coordinates and display the landmarks at their corresponding locations. As the coordinates are retrieved, euclidean distances, using the same Equation \ref{equationdist} between the specific points can be computed. These distances, as explained before, correspond to the anthropometric measurements established by the HUTUBS dataset for the pinna and are mentioned in Table \ref{Hutubs features}. 
\medskip

\begin{figure}[!htbp]
     \centering
     \includegraphics[width=8cm]{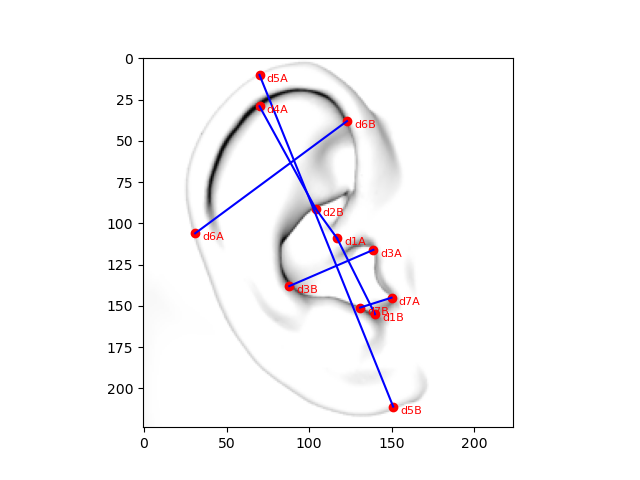}
     \caption{Subject 2 left ear annotated with landmarks (red) and distances (blue) in Python}
    \label{subject2pythLD}
\end{figure}

The distances can be plotted on the image with the specific landmarks for visualisation purposes. Figure \ref{subject2pythLD} gives a representation of the distances plotted from the appropriate landmarks for subject number 2 in the HUTUBS database. Additional images are shown in Figure \ref{hu_12}.
\medskip

The 7 distances being computed on the 224x224 pixels image using the euclidean distances equation, they are normalized with the diagonal of the 224x224 square which is rounded to 316 pixels as it is the maximum distance that could be obtained with this image format giving distances varying between 0 and 1. The normalized distances are then stored in an Excel file for each ear. 
\medskip

To summarize, the landmarks on the HUTUBS subject ear images have been positioned, the normalized distances in pixels have been retrieved and stored in an excel file, therefore the conversion factors can be computed.  For each distance of each ear, the measurements in centimetres provided by HUTUBS is divided by the normalized distance in pixel which gives a conversion factor between a normalized distance in pixel and in centimetres. A factor is computed for each distances and for each ear, it is stored in an Excel file and then an average factor per distance can be computed over the 116 ears.
\medskip
\newpage
For the $i^{th}$ ear, the $j^{th}$ distance $C_{ij}$ in centimetres can be divided by the $j^{th}$ distance $P_{ij}$ in pixels to obtain the conversion factor $f_{ij}$. 

\begin{equation}
    f_{ij} = \frac{C_{ij}}{P_{ij}} 
\end{equation}
For i = 1,...,116 and j = 1,...,7.
\medskip
The average factor per distance $F_j$ can be computed by dividing the sum of every $f_{ij}$ ear conversion factor per 116. 

\begin{equation}
    F_j = \frac{\sum_{i=1}^{116}f_{ij}}{116}
\end{equation}
For j = 1,...,7.
\medskip
Each distance has its conversion factor to convert a normalized distance in pixel to a distance in centimetres. The conversion factor average on the 116 ears are listed in Table \ref{conversion factor}. 

\begin{table}[!htbp]
\centering
\begin{tabular}{ |c|c|c|}
\hline
         Conversion factor   & Measurements & Value [cm/pixel]\\
    \hline

     $F_1$ & cavum concha height & 10.129765\\
    \hline
     
     $F_2$ & cymba concha height & 13.442287\\
    \hline
    
     $F_3$ & cavum concha width & 11.625544\\
    \hline
   
      $F_4$ & fossa height & 9.539581\\
    \hline
  
    $F_5$ & pinna height & 8.621989\\
    \hline
     
     $F_6$ & pinna width & 11.824525\\
    \hline
 
      $F_7$ & intertragal incisure width & 10.532984\\
    \hline
    
      $F_A$ & Overall Average  & 10.313797\\
    \hline
    
\end{tabular}
    \caption{Seven measurements with their corresponding conversion factor}
    \label{conversion factor}
\end{table}

As an observation, the conversion factors do not vary much and oscillate around 10 which gives an interesting overall result. To validate the conversion factors, the ear images retrieved from the 3D meshes are given as inputs to the neural network with the aim to find the 7 distances in pixels, convert them into centimetres and compare the result to the actual distances provided. 
\medskip

Unfortunately, the result of the CNN does not correctly position the landmarks on the image. The assumed reason for this could be that the image is a black and white image of the mesh and does not resemble to the images of the training dataset as they have no colour and the human skin contrast is not present. Considering this factor, the CNN is not effective on this type of image. Additionally, as mentioned earlier, the position of the landmarks which serve as references for the computation of distances in pixels are assumed and not given accurately by HUTUBS. This latter finding introduces uncertainty and irregularities in the validity of the conversion factor. 
\medskip

As a result, the conversion factors cannot be validated but only used as a short-term solution. This approach was an interesting attempt to retrieve distances in 224x224 pixels which do not contain any reference of distance. The most rational and ideal solution would be to have ear images which contain a distance reference. This approach will be explained below in Section \ref{pix} along with further developments. Nevertheless, the conversion factors can temporarily be used to obtain distances in centimetres and find a best match profile in the HUTUBS HRTFs database.

\section{HUTUBS database best match}

This final section of the HRTF individualization process aims to use the distances in centimetres obtained with the conversion factors to find the profile with the closest measurements in an existing Database. The goal is to find a correspondent ear which matches as close as possible to the concerned subject ear characteristics. HUTUBS database \cite{hutub} is one of the only databases to provide anthropometric measurements as well as corresponding acoustically measured HRTFs. Indeed, they provide 58 subjects anthropometric measurements along with their HRTFs. The 7 chosen distances are isolated from the original file and stored in an excel file for each subject of the database and for each ear. A set of 116 ears measurements are thus available to find a best match. 
\medskip

Pelzer et al. \cite{Pelzer} stated that the HUTUBS database is sufficiently large enough to provide well-matching nonindividual HRTFs, in many cases, and sufficiently diverse enough to produce large differences between individual and nonindividual HRTFs, which are required for recommendation-based HRTF individualization. The vector of 7 distances obtained from the ear picture and converted in centimeters can be compared to the vectors of distances provided by the HUTUBS database. 
\medskip

Let's call $d_{ear} = [d_1, d_2, \ldots, d_7] $ the obtained vector of 7 distances and $D_{HUTUBS} = [D_1, D_2, \ldots, D_{116}] $, the database of distances where each $D_i = [d_{i1}, d_{i2}, \ldots, d_{i7}] $ with $i = 1,2,...,116$. The Euclidean distance between two vectors a and b is given by:
\begin{equation}
    Euc_{(a, b)} = \sqrt{\sum_{i=1}^{n} (a_i - b_i)^2} 
\end{equation}

For the given vector $d_{ear} $ and each vector $D_i $ in the database, the Euclidean distance is calculated as:

\begin{equation}
    Euc_i = \sqrt{ \sum_{j=1}^{7} (d_{ear_j} - d_{ij})^2}
\end{equation}
For i = 1,...,116.
\medskip

The code computes the Euclidean distances for all vectors in the database. The vector $ D_i $ that corresponds to the smallest value of $Euc_i $ is considered the best match, as it has the smallest Euclidean distance to the given vector $d_{ear} $. The corresponding number of the ear of the best match is retrieved allowing the number of the subject whose ear best matches the ear given in the picture to be identified. The dataset contains the HRTFs in SOFA format so the result match HRTF can be used with its SOFA format for the application needed. 
\medskip

Due to the non-validity of the conversion factor, the best match HRTF set will not be evaluated as it does not correctly represent the anthropometric features of the subjects ear. The process functions but it is not valid for a proper HRTF individualization in the present state. Further developments will allow for better validity of the process and then an evaluation of the matched HRTF set but it will not be explored in this work. 
\medskip

This concludes the automated HRTFs individualization process from a picture of a subject's ear to the HRTF set in SOFA format ready to use with minimal effort or intervention from the user. The entirety of the process works effectively and every part can be optimized, better designed and explored further. These modifications and improvements are discussed in the next section for further developments.

\chapter{Further developments} \label{further}

Following  a comprehensive review and explanation of the different steps of the HRTF individualization process, many improvements can be explored with the aim of improving each stage which will bring more validity to the process. Indeed, as it is the initial implementation of this type of individualization process, many  thoughts and ideas have emerged in order to create a more efficient process and bring more efficiency to each part. Specific elements of the process will be addressed in this section such as the neural network dataset, the CNN architecture, the conversion factor and the HRTF database. 

\section{The neural network dataset} 

The dataset currently contains 3000 training ear images and 630 test ear images which have been collected on Google Image. 92\% of the pictures have not been taken directly in front of the ear, the majority of the original images, prior to being re-framed and distorted, are photographs of celebrities which include their entire body. The ear images in the database are low in resolution due to the fact that the ear is not the primary focus of the original picture and the ears represent a small portion of the full photography resolution, hence the need of re-framing. 
\medskip

The ear images do not contain a measurement reference adjacent to the ear which makes it hardly possible to accurately retrieve the exact measurement in centimetres of the anthropometric features chosen for the ear. The pictures are also resized in order to have the same size of 224x224 pixels, this means that the image of the ear which is fed into the neural networks is in fact a distorted representation. 
\medskip

Furthermore, for each ear picture there are 55 landmarks, as explained in Section \ref{12}, and only 12 points are required to represent 7 distances which are relevant as anthropometric measurements within the HRTF individualization process. Limiting the selection to 12 points as landmark in the image and 2 for the reference distance, would means that the output layer size of the neural network will be reduced to 28 instead of 110. The reduction of the output layer of the CNN while maintaining the same network architecture could lead to significant changes in the network performance. 
\medskip

For all these reasons a new database is absolutely essential for a coherent learning process which can lead to several improvements for the following steps of the process. The dataset would be designed accordingly to the following specific criteria.
\medskip

Ideally, the amount of subjects would be increased for greater diversity constituting a larger quantity of ear images. It is imperative that these pictures are taken facing the ear with sufficient quality to be cropped in a square format containing the entire ear and the distance reference. The reference distance could be a specific pattern with noticeable reference points which the neural network could learn in order to recognise and position the points on the picture precisely. Once the pictures have been acquired with the correct specifications, a manual work would have to be done to mark every 12 points on each ear plus 2 points for the distance reference points. Once all the landmarks are associated for each image, a data augmentation step can be realized to increase the training and test data of the CNN. 
\medskip

In that context, the CNN would be learning on a greater number of undistorted, high resolution ear images which are facing the ear and annotated with 12 landmarks plus 2 for the reference distances. 

\section{The CNN architecture}

The CNN structure can be studied and improved for greater performance and general faster processing. Through careful analysis of the dataset and the concerned task requirements, adjustments to the CNN's depth, width, and layer configurations could be made. This step might involves fine-tuning hyperparameters, such as kernel size, filter count, and pooling strategies, to adapt to the new dataset. As mentioned above, the output layer size would be reduced to 28, instead of 110, which could have an impact the CNN performance. By iteratively experimenting with modifications, it would be possible to achieve a more efficient architecture tailored to the characteristics of the updated dataset, resulting in improved accuracy and generalization. The new network performance would have to be tested and validated through meticulous assessments. 

\section{The pixels to centimetres conversion} \label{pix}

By introducing an established physical distance within the image, such as a ruler or a reference object with a precisely measured length between two visible points, it is possible to establish a reliable mapping between pixel distances and real-world measurements in centimetres. This reference distance pattern would function as a calibration mechanism, allowing the system to accurately convert pixel coordinates to centimeters. This approach not only enhances the precision of measurements but also mitigates potential distortions caused by variations in image resolution, camera angles and settings. As a result, incorporating a reference distance pattern serves as a crucial step in ensuring the accuracy and consistency of geometric measurements within images, ultimately leading to improved validity of the process.

\section{The HRTF database}

Despite Pelzer et al. \cite{Pelzer} stating that the HUTUBS HRTF database is sufficiently large enough to provide well-matching nonindividual HRTFs. A larger database of SOFA HRTF would increase the diversity of ear anthropometry and anatomy features. A larger amount of measurement profiles would allow for a smaller euclidean distance between vectors leading to a closer match of the 7 distances vector. Ideally, he HRTF would have to be measured acoustically or simulated and associated with the subject's ear picture with a reference distance for anthropometric distances computation. 

\section{Influences of other anthropometric measurements}

Xu et al. \cite{xu1} focused their work on finding the morphological factors which influence HRTFs by analyzing the CIPIC database using Principal Component Analysis (PCA). They conclude that more measurements of the ear exhibit a substantial correlation with HRTF magnitudes in higher frequencies, while the torso and head demonstrate a significant correlation with HRTFs in lower frequencies due to the diffraction acoustic effect. Greater horizontal measurements display a correlation with HRTF magnitudes in lower frequencies compared to higher frequencies, whereas vertical and angular measurements exhibit an opposite pattern. Their correlation analysis state that three components which mainly represent the measurements of the head, the shoulder and ear angles have significant influences on the magnitudes of HRTFs. 
\medskip

This indicates that more anthropometric measurements have to be considered when providing an HRTF individualization. Therefore, a wider selection of anthropometric features must play a role in the selection of a specific set of HRTF for an individual. These features can be also retrieved from images using convolutional neural networks. A dataset would have to be designed to train a neural network to predict landmarks on other views of the subjects. It is an interesting path to follow in order to provide higher tailored HRTF to someone anatomy.

\chapter*{Conclusion}

The objective of this Master Thesis was to establish a solid understanding of spatial sound, HRTF and how artificial intelligence could play a role in innovative approaches. A review of what research and studies have brought along the years since 1970s to the evolution of immersive audio 3D experiences was conducted and led to the understanding in the demand for easier access of individualized acoustic filters using artificial intelligence technologies. 
\medskip

The objective was to develop the initial stages of an automated HRTF Individualization using neural networks and available databases. The goal was to design a process with the easiest way for the user to obtain his individualized set HRTFs. Every step of the process is though for as little effort as possible from the user side. The primary goal was not to tailor perfectly the subject but to develop a process with various stages which could attribute a set of HRTFs for an individual based on the ear image using anthropometric features.
\medskip

A process of HRTF individualization using a convolutional network was developed and capable of positioning landmarks on ear images, selecting the most relevant ones, computing anthropometric distances, converting them into centimetres and to find a best match set of HRTFs in an existing HRTFs database. This process requires a detailed image of the subjects ear in order to obtain a specific individualized set of HRTFs which are tailored to the unique profile of the ear. 
\medskip

During this work, several limitations were identified which appeared to be detrimental to the validity of specific elements used in the process. Individual elements of the process can be optimized and tailored to the desired task. The conceptual scope of each stage has been verified and substantiated to function correctly. The goal was to demonstrate that this HRTF individualisation pipeline is valid and ready for greater investment and development. 
\medskip

Additional research needs to be considered in order to develop a full product of this prototype concept. Several designs and research developments have to be explored further to create a perfectly functioning process of HRTF individualization. Spatial audio and HRTFs have the scope to introduce many innovations within its specialized field. 
\medskip

Sound spatialization is a ground breaking technology which will have a far reaching impact for individuals, the scientific community and the audio industry.

\appendix
\chapter{Appendix A}
\section{Deep learning and CNN} \label{deeplearning}

\begin{figure}[!htbp]
     \centering
     \includegraphics[width=8cm]{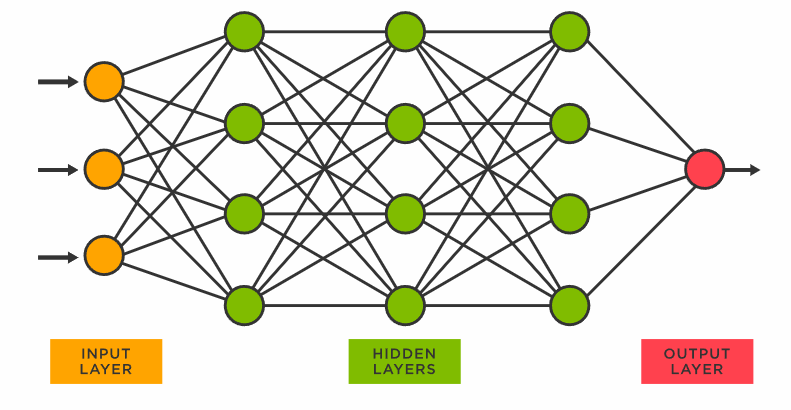}
     \caption{Artificial Neural Network structure}
    \label{NN}
\end{figure}

Deep Learning is a subset of Machine Learning, which includes all the techniques and algorithms that enable a machine to learn and improve autonomously. It is also known as artificial intelligence. Deep Learning uses artificial neural networks inspired, as the name suggests, by the neurons in the human brain. These artificial neural networks consist of neurons (circles on Figure \ref{NN}\footnote{Tibo.com : What is a Neural Network ? Consulted in June 2023. \url{https://www.tibco.com/reference-center/what-is-a-neural-network}}) resembled in several layers of neurons. In a fully connected layer, each neuron in the layer is connected to every neuron in the previous layer and every neuron in the next layer, however, not all layers in a network are necessarily fully connected. A neural network is composed of at least two layers, an input layer through which the data is inserted and an output layer giving the predictions of the network. These two layers are separated by a set of intermediate layers known as hidden layers. The more hidden layers a network has, the deeper it is, known as a deep neural network. Deep Learning therefore covers all machine learning techniques that use deep neural networks.

In order to learn, a neural network must be trained on a large database. The training of such a network is based on two principles: 'feedforward' (from the input layer to the output layer) and 'backpropagation' (from the output layer to the output layer).
\medskip

Feedforward describes the way in which information is propagated within the network from the input layer to the output layer. To do this, each neuron corresponds to a value and each connection to a weight. The value of a neuron depends on the values of the neurons in the previous layer and the weight of each connection with these neurons by the following formula : 

\begin{equation}
    f(x) = \sigma(x_1w_{j1} +x_2w_{j2}+...+x_nw_{jn}+b_j)
\end{equation}

Knowing that the $x_i$ correspond to the values of the previous neurons (the inputs) and the $w_{ji}$ to the weights, the value of a neuron therefore corresponds to a weighted sum of the different inputs where the weight reflects the influence of a certain input on the output and therefore the value of a neuron. The bias $b_j$ is added to this weighted sum. Bias can take a positive or negative value to determine when the neuron's value starts to be significant. In this way, a bias is associated with each neuron in the network. Finally, an activation function $\sigma$, which is a mathematical function used to introduce nonlinearity into the network. There are different types of activation function, the most commonly used being ReLU, sigmoid, and tanh.
\medskip

It is through this principle that the 'Feedforward' allows information to propagate within the network until its obtains the output data. In the case of a supervised problem, each data is accompanied by a label, also called an annotation, which corresponds to the desired result for that data. For instance, it could be positions of landmarks for an image prediction problem. These labels enable the use of a cost or loss function that estimates the error of the network's output value compared to the actual value. There are different cost functions, such as Mean Squared Error (MSE). The choice of the cost function depends, in particular, on the application.

\begin{figure}[!htbp]
     \centering
     \includegraphics[width=8cm]{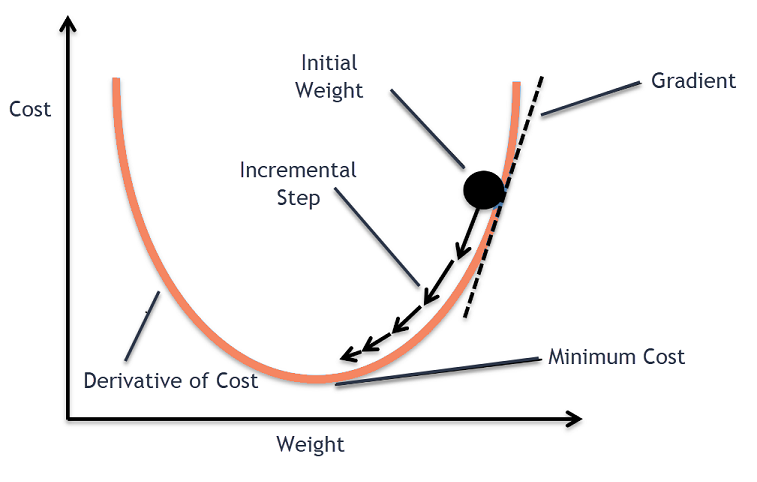}
     \caption{Gradient Descent}
    \label{gradient}
\end{figure}

Now, let's explain how the neural network works. This is where 'Backpropagation' comes into play. To improve the performance of the neural network, the goal is to find the minimum of the cost function, making it an optimization problem. Backpropagation involves updating the network's weights during training to enhance its performance. However, this update is not performed randomly. To optimize the performance, an "Optimizer" is used, which is an algorithm that aims to change certain parameters of the network, such as weights and learning rate, to reduce losses. Different optimizers, such as Stochastic Gradient Descent (SGD) (see Figure \ref{gradient}\footnote{Analyticsvidhya.com : How Does the Gradient Descent Algorithm Work in Machine Learning ? Consulted in June 2023. \url{https://www.analyticsvidhya.com/blog/2020/10/how-does-the-gradient-descent-algorithm-work-in-machine-learning/}}) and Adam, exist, but all utilize the principle of gradient descent.
\medskip

To illustrate this principle, consider the simplest case with only one parameter (weight). The gradient descent algorithm starts with a random value for the parameter and calculates the gradient, which is a vector of partial derivatives indicating the direction to reach the function's minimum. The algorithm then advances the parameter's value in that direction using a certain step called the learning rate. The gradient is recalculated to move the parameter again, and this process continues until the global minimum of the function is found. It is an iterative method. In reality, a neural network consists of thousands of parameters, making the cost function multidimensional. Therefore, a large dataset is necessary to gradually find the global minimum. With this in mind, 'Backpropagation' aims to calculate the gradients of the internal layers starting from the error at the output, in order to update the weights and biases of each layer.
\medskip

To train a neural network, it is necessary to split the dataset into three sets. One is the training set used to train the model, another is the validation set used to evaluate the network's performance on unseen data during training, and finally, the test set is used to assess the final performance of the neural network after training. When all the training data has been passed through the network once during training, it completes one epoch. The number of epochs to be performed during training is a parameter that needs to be determined before starting the process. The number of data samples sent to the network before updating its weights is referred to as a batch.

\subsection{Convolutional Neural Network (CNN)} \label{CNN}

\begin{figure}[!htbp]
     \centering
     \includegraphics[width=10cm]{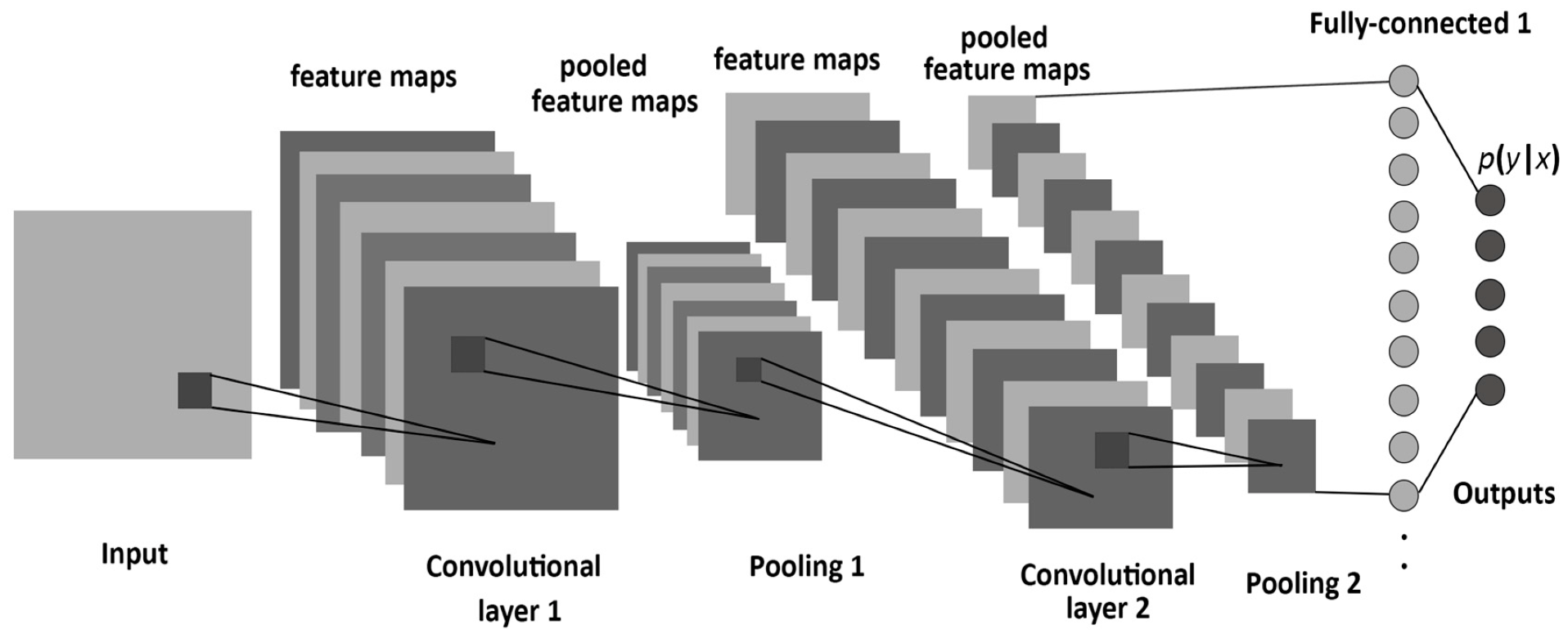}
     \caption{Convolutional Neural Network architecture}
    \label{cnnarch}
\end{figure}

A Convolutional Neural Network (CNN) is a type of neural network used to extract high-level features from images. It consists of three types of layers: convolutional layers, pooling layers, and fully-connected layers. Figure \ref{cnnarch}\footnote{Towardsdatascience.com : How To Teach A Computer To See  With Convolutional Neural Networks ? Consulted in June 2023. \url{https://towardsdatascience.com/how-to-teach-a computer-to-see-with-convolutional-neural-networks-96c120827cd1}} shows a traditional convolutional neural network.

\medskip

The convolutional layers represent the majority of layers in a CNN and play a crucial role in feature extraction. They involve a higher number of layers compared to the fully connected layers, which are used mainly for classification based on these features. Since an image is essentially a matrix where each value corresponds to a pixel, a convolutional layer performs convolutions between this matrix and a set of other matrices called "kernels" or filters to obtain "activation maps" or "feature maps." All the kernels have the same size, which is smaller than that of the image. As the kernel moves across the entire image, it performs a convolution operation with the corresponding matrix of the image area, resulting in an activation map. Each kernel has different matrix values, so the activation maps obtained are all different. This method allows the CNN to extract image features. The initial convolutional layers of the CNN extract low-level features and progress towards extracting high-level features in later layers. During training, the values of these kernels are modified to extract features more effectively, and these matrix values can be considered as the weights of the neural network. After each convolutional layer, a ReLU (Rectified Linear Unit) activation function is typically used to introduce non-linearity into the activation maps.
\medskip

In CNNs, several pooling layers are typically used to reduce the spatial size of activation maps, while their number increases along the network. This helps to decrease the computational requirements. Finally, a CNN usually concludes with fully connected layers, which are more traditional neuron layers as described in the previous section. This enables the use of high-level features obtained earlier to determine the image class in a classification problem.

\section{The CNN model layers} \label{layer}

This section provides a detailed exploration of the different layers comprising the Convolutional Neural Network (CNN) used in \ref{neural network}. 
\medskip

Here is the detailed architecture of the proposed CNN : 

\begin{figure}[!htbp]
     \centering
     \includegraphics[width=12cm]{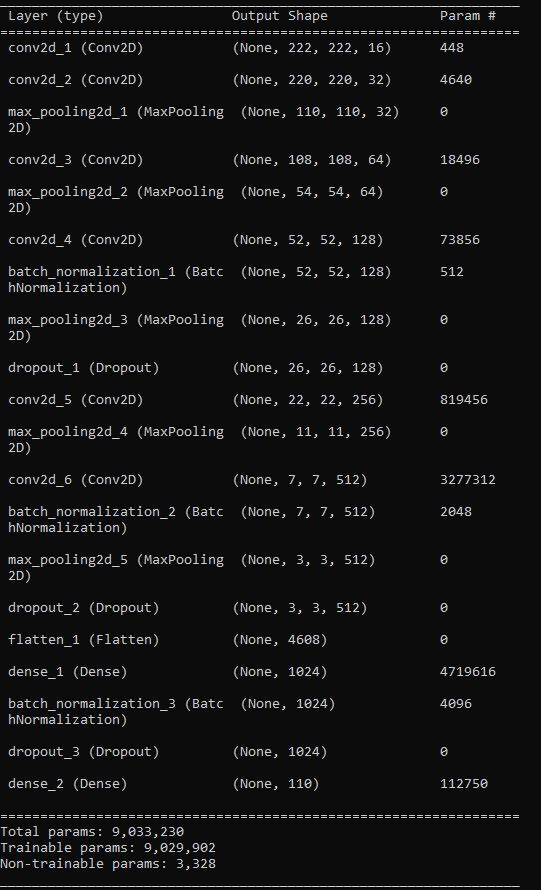}
     \caption{Convolutional Neural Network model architecture}
    \label{model}
\end{figure}

\begin{itemize}
    \item $Conv2D_1$ layer has 16 filters with a size of (3, 3). It takes the input shape of (224, 224, 3) and produces an output shape of (222, 222, 16). It has a total of 448 parameters.
    \item $Conv2D_2$ layer has 32 filters with a size of (3, 3). It takes the previous layer's output shape as input and produces an output shape of (220, 220, 32). It has a total of 4,640 parameters.
    \item $MaxPooling2D_1$ layer performs max pooling with a pool size of (2, 2) on the previous layer's output. It reduces the spatial dimensions by half, resulting in an output shape of (110, 110, 32).
    \item $Conv2D_3$ layer has 64 filters with a size of (3, 3). It takes the previous layer's output shape as input and produces an output shape of (108, 108, 64). It has a total of 18,496 parameters.
    \item $MaxPooling2D_2$ layer performs max pooling with a pool size of (2, 2) on the previous layer's output. It reduces the spatial dimensions by half, resulting in an output shape of (54, 54, 64).
    \item $Conv2D_4$ layer has 128 filters with a size of (3, 3). It takes the previous layer's output shape as input and produces an output shape of (52, 52, 128). It has a total of 73,856 parameters.
    \item $BatchNormalization_1$ layer normalizes the activations of the previous layer, helping in faster training and improved generalization. It has a total of 512 parameters.
    \item $MaxPooling2D_3$ layer performs max pooling with a pool size of (2, 2) on the previous layer's output. It reduces the spatial dimensions by half, resulting in an output shape of (26, 26, 128).
    \item $Dropout_1$ layer randomly deactivates 30 \% of the neurons in the previous layer during training, helping to prevent overfitting.
    \item $Conv2D_5$ layer has 256 filters with a size of (5, 5). It takes the previous layer's output shape as input and produces an output shape of (22, 22, 256). It has a total of 819,456 parameters.
    \item $MaxPooling2D_4$ layer performs max pooling with a pool size of (2, 2) on the previous layer's output. It reduces the spatial dimensions by half, resulting in an output shape of (11, 11, 256).
    \item $Conv2D_6$ layer has 512 filters with a size of (5, 5). It takes the previous layer's output shape as input and produces an output shape of (7, 7, 512). It has a total of 3,277,312 parameters.
    \item $BatchNormalization_2$ layer normalizes the activations of the previous layer. It has a total of 2,048 parameters.
    \item $MaxPooling2D_5$ layer performs max pooling with a pool size of (2, 2) on the previous layer's output. It reduces the spatial dimensions by half, resulting in an output shape of (3, 3, 512).
    \item $Dropout_2$ layer randomly deactivates 50\% of the neurons in the previous layer during training.
    \item $Flatten_1$ layer flattens the previous layer's output into a 1-dimensional vector with a shape of (4608).
    \item $Dense_1$ layer has 1024 neurons with a ReLU activation function. It takes the flattened input and produces an output shape of (1024). It has a total of 4,719,616 parameters.
    \item $BatchNormalization_3$ layer normalizes the activations of the previous layer. It has a total of 4,096 parameters.
    \item $Dropout_3$ layer randomly deactivates 70\% of the neurons in the previous layer during training.
    \item $Dense_2$ layer has 110 neurons, representing the output dimension for the classification task. It does not have an activation function specified.
\end{itemize}

\chapter{Appendix B}
\section{Related figures}
Here are more figures related to this work. 

\begin{figure}[!htbp]
     \centering
     \includegraphics[width=13cm]{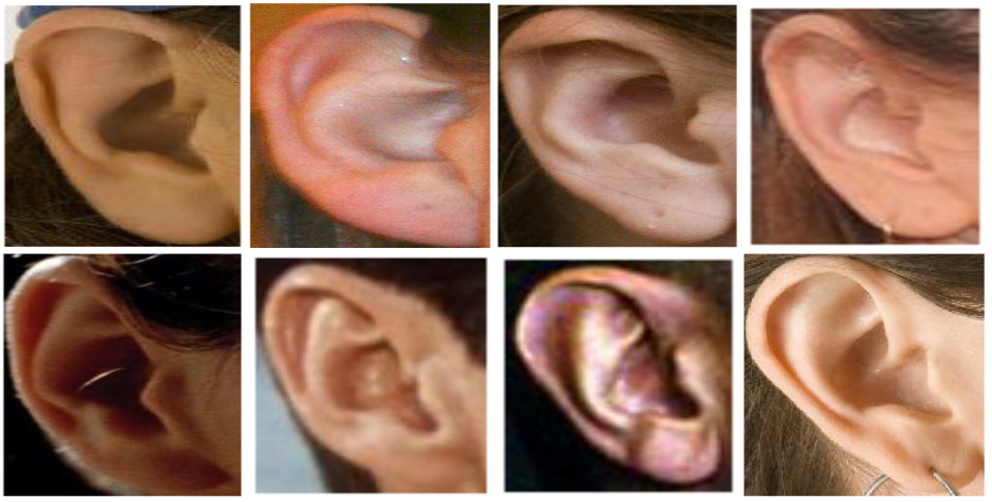}
     \caption{8 example images of ear pictures of the I-BUG dataset}
    \label{8}
\end{figure}

\begin{figure}[!htbp]
     \centering
     \includegraphics[width=13cm]{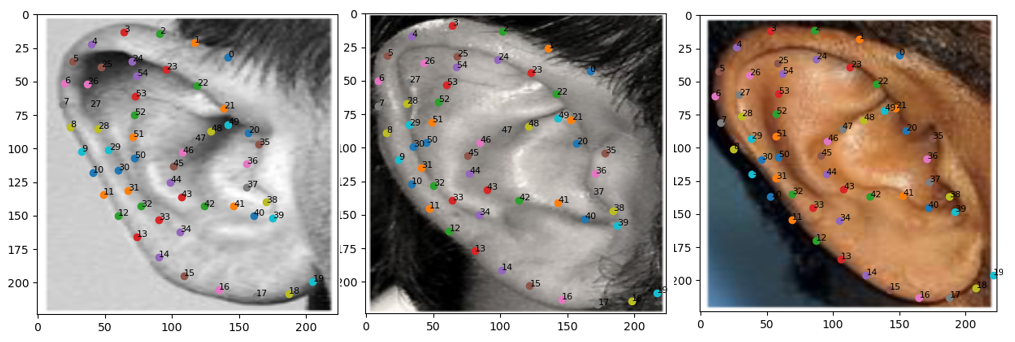}
     \caption{3 example images of the output CNN result with the 55 predicted landmarks in color dots with their corresponding label}
    \label{3_55}
\end{figure}

\begin{figure}[!htbp]
     \centering
     \includegraphics[width=13cm]{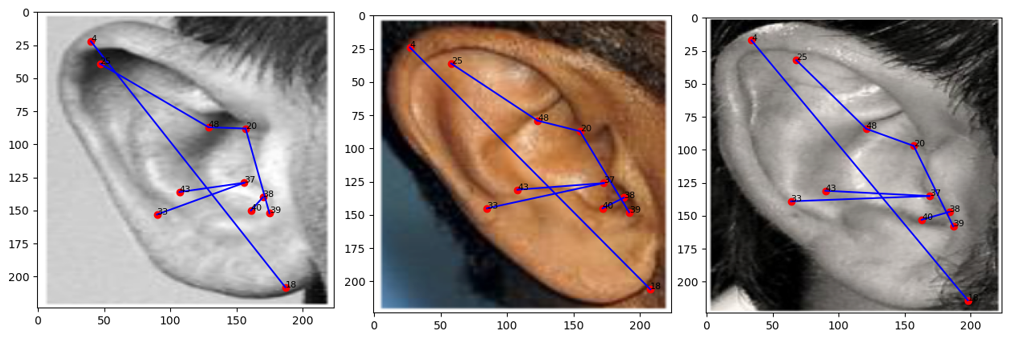}
     \caption{3 example images with the chosen predicted landmarks in red dots and connected with their corresponding distances in blue}
    \label{3_12}
\end{figure}

\begin{figure}[!htbp]
     \centering
     \includegraphics[width=13cm]{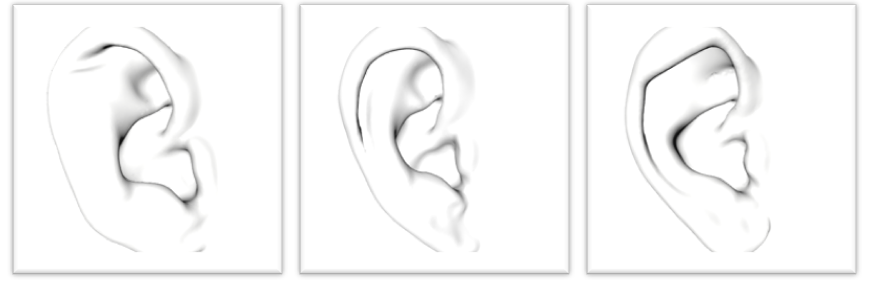}
     \caption{3 example images of the HUTUBS ears retrieved with Python from 3D head meshes}
    \label{hu_cadre}
\end{figure}

\begin{figure}[!htbp]
     \centering
     \includegraphics[width=13cm]{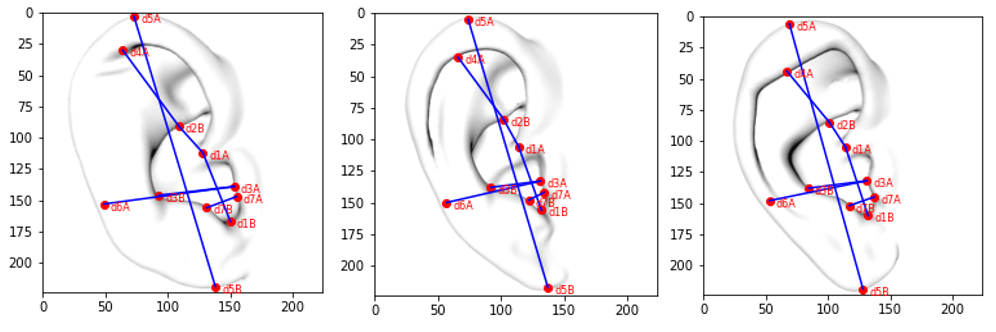}
     \caption{3 example images of the HUTUBS ears with the chosen landmarks in red dots and connected with their corresponding distances in blue}
    \label{hu_12}
\end{figure}

\begin{figure}[!htbp]
     \centering
     \includegraphics[width=13cm]{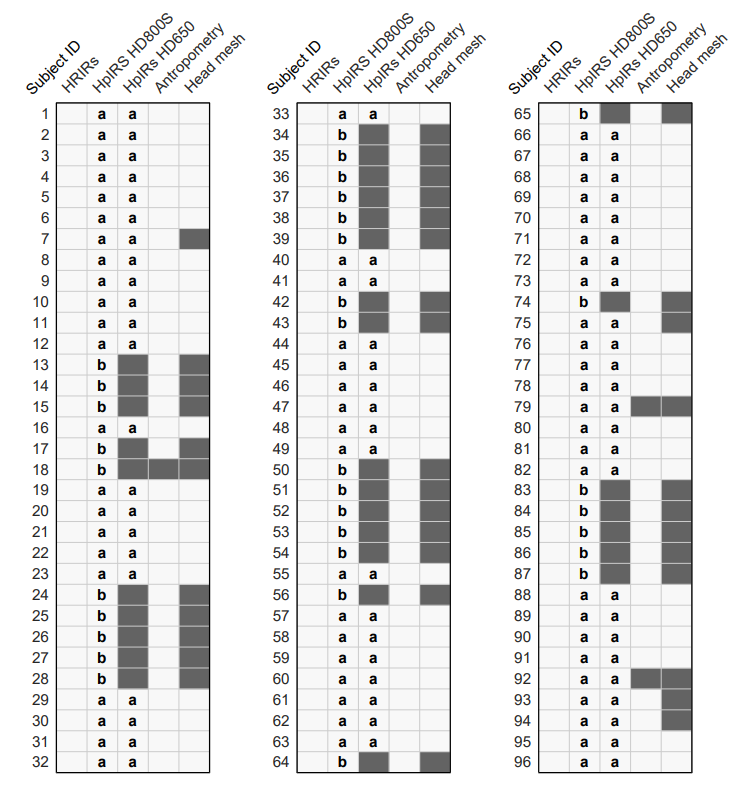}
     \caption{The HUTUBS overview of available data for each subject \cite{hutub}. a and b letters represent the two different Sennheiser HD800S headphones that were used during the acquisition of the database. If the box is grey, the data is not available}
    \label{hu_over}
\end{figure}

\begin{figure}[!htbp]
     \centering
     \includegraphics[width=14cm]{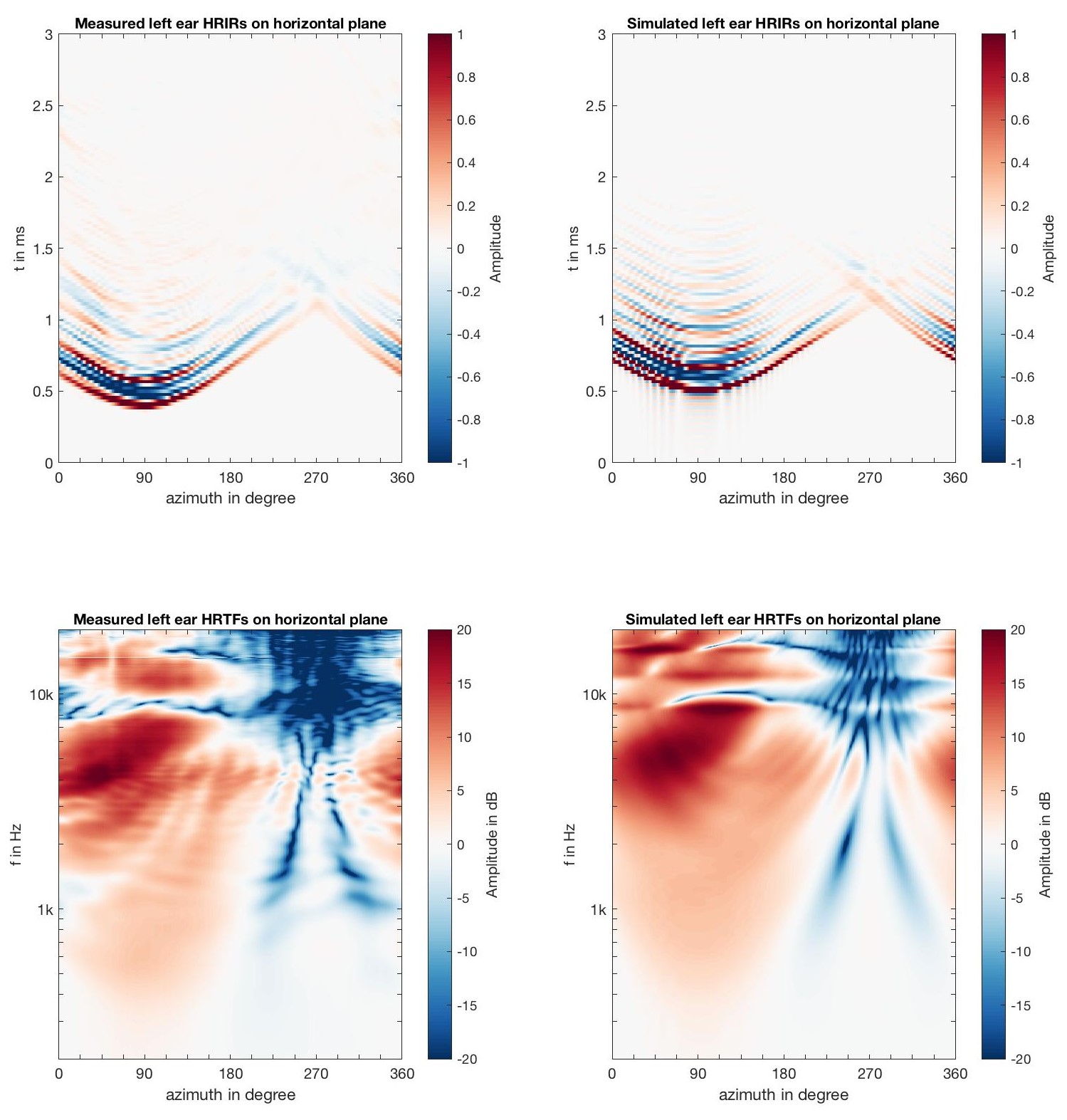}
     \caption{HUTUBS figures of the subject 2 HRTF on horizontal plane \cite{hutub}}
    \label{hu_sub2}
\end{figure}

\end{document}